\documentclass[reqno,draft,10pt]{amsart}

\usepackage{verbatim}

\theoremstyle{plain}
\newtheorem{lemma}{Lemma}
\newtheorem{theorem}[lemma]{Theorem}
\newtheorem{corollary}[lemma]{Corollary}

\newtheorem{proposition}[lemma]{Proposition}
\newtheorem{definition}{Definition}

\theoremstyle{remark}
\newtheorem{remark}{Remark}

\newcommand*  {\N} {{\mathbb N}}

\newcommand*  {\R} {{\mathbb R}}

\newcommand*  {\C} {{\mathbb C}}

\newcommand*{\norm}[3][{\vphantom 1}]{\lVert #2 \rVert_{#3}^{#1}}

\newcommand*{\abs}[1]{\lvert #1 \rvert}
\newcommand*{\babs}[1]{\bigg \lvert #1 \bigg \rvert}

\newcommand*{\ang}[1]{\left\langle #1 \right\rangle}

\newcommand*{\Space}[3]{{#1}_{{\vphantom 1}#2}^{{\vphantom +}#3}}

\renewcommand*{\l} [2]{\Space{L}{#1}{#2}}

\newcommand{\nm}   [1]{\norm{#1}{}{}}
\newcommand{\ddt}     {\frac{d}{dt}}
\newcommand{\pdt}  [1]{\frac{d #1}{d t}}
\newcommand{\half}    {\frac{1}{2}}

\begin{document}

\title[Analytic Study of Shell Models of Turbulence]%
{Analytic Study of Shell Models of Turbulence}

\date{May 31, 2006}
\thanks{{\bf To appear in:} {\it Physica D}}

\author[P. Constantin]{Peter Constantin}
\address[P. Constantin]%
{Department of Mathematics \\
 The University of Chicago \\
 Chicago, IL 60637 \\
 USA}
\email{const@math.uchicago.edu}

\author[B. Levant]{Boris Levant}
\address[B. Levant]%
{Department of Computer Science and Applied Mathematics \\
 Weizmann Institute of Science \\
 Rehovot, 76100 \\
 Israel}
\email{boris.levant@weizmann.ac.il}

\author[E. S. Titi]{Edriss S. Titi}
\address[E. S. Titi]%
{Department of Mathematics and
 Department of Mechanical and Aerospace Engineering \\
 University of California \\
 Irvine, CA 92697 \\
 USA \\
 Also, Department of Computer Science and Applied Mathematics \\
 Weizmann Institute of Science \\
 Rehovot, 76100 \\
 Israel}
\email{etiti@math.uci.edu}

\begin{abstract}
In this paper we study analytically the viscous  `sabra' shell model
of energy turbulent cascade. We prove the global regularity of
solutions and show that the shell model has finitely many asymptotic
degrees of freedom, specifically: a finite dimensional global
attractor and globally invariant inertial manifolds. Moreover, we
establish the existence of exponentially decaying energy dissipation
range for the sufficiently smooth forcing.
\end{abstract}

\maketitle

\medskip

\medskip


\tableofcontents
\newpage

\section{Introduction}

Shell models of turbulence have attracted  interest as useful
phenomenological models that retain certain features of the
Navier-Stokes equations (NSE). Their central computational advantage
is the parameterization of the fluctuation of a turbulent field in
each octave of wave numbers $\lambda^{n} < \abs{k_n} \le
\lambda^{n+1}$ by very few representative variables. The octaves of
wave numbers are called shells and the variables retained are called
shell variables. Like in the Fourier representation of NSE, the time
evolution of the shell variables is governed by an infinite system
of coupled ordinary differential equations with quadratic
nonlinearities, with forcing applied to the large scales and viscous
dissipation effecting the smaller ones. Because of the very reduced
number of interactions in each octave of wave numbers, the shell
models are drastic modification of the original NSE in  Fourier
space.

The main object of this work is the ``sabra'' shell model of
turbulence, which was introduced in \cite{LP98}. For other shell
models see \cite{BJPV98}, \cite{Gl73}, \cite{Fr95}, \cite{OY89}. A
recent review of the subject emphasizing  the applications of the
shell models to the study of the energy-cascade mechanism in
turbulence can be found in \cite{Bi03}. It is worth noting that the
results of this article apply equally well to the well-known
Gledzer-Okhitani-Yamada (GOY) shell model, introduced in
\cite{OY89}. The main difference between shell models and the NSE
written in Fourier space is the fact that the shell models contain
only local interaction between the modes. We like to point out that
our analysis could be performed also in the presence of the long,
but finite, range interactions.

The sabra shell model of turbulence describes the evolution of
complex Fourier-like components of a scalar velocity field denoted
by $u_n$. The associated one-dimensional wavenumbers are denoted by
$k_n$, where the discrete index $n$ is referred to as the ``shell
index''. The equations of motion of the sabra shell model of
turbulence have the following form
\begin{equation} \label{eq_sabra}
\frac{d u_n}{d t} = i (a k_{n+1} u_{n+2} u_{n+1}^* + b k_{n}
u_{n+1} u_{n-1}^* - c k_{n-1} u_{n-1} u_{n-2}) - \nu k_n^2 u_n +
f_n,
\end{equation}
for $n = 1, 2, 3, \dots$, and the boundary conditions are $u_{-1}
= u_{0} = 0$. The wave numbers $k_n$ are taken to be
\begin{equation} \label{eq_freq}
k_n = k_0 \lambda^n,
\end{equation}
with $\lambda  > 1$ being the shell spacing parameter, and $k_0
> 0$. Although the equation does not capture any geometry, we will
consider $L = k_0^{-1}$ as a fixed typical length scale of the
model. In an analogy to the Navier-Stokes equations $\nu > 0$
represents a kinematic viscosity and $f_n$ are the Fourier
components of the forcing.

The three parameters of the model $a, b$ and $c$ are real. In
order for the sabra shell model to be a system of the hydrodynamic
type we require that in the inviscid ($\nu = 0$) and unforced
($f_n = 0$, $n = 1, 2, 3, \dots$) case the model will have at
least one quadratic invariant. Requiring conservation of the energy
\begin{equation} \label{eq_energy}
\mathrm{E} = \sum_{n=1}^\infty \abs{u_n}^2
\end{equation}
leads to the following relation between the
parameters of the model, which we will refer as an energy
conservation condition
\begin{equation} \label{eq_energy_cons_assumption}
a + b + c = 0.
\end{equation}
Moreover, in the inviscid and unforced case the model possesses
another quadratic invariant
\begin{equation} \label{eq_helicity}
\mathrm{W} = \sum_{n=1}^\infty \bigg( \frac{a}{c} \bigg)^n
\abs{u_n}^2.
\end{equation}

The physically relevant range of parameters is $\abs{a/c}
< 1$ (see \cite{LP98} for details). For $-1 < \frac{a}{c} < 0$
the quantity $\mathrm{W}$ is not sign-definite and therefore it is
common to associate it with the helicity -- in an analogy to the
$3$D turbulence. In that regime we can rewrite relation
(\ref{eq_helicity}) in the form
\begin{equation} \label{eq_helicity2}
\mathrm{W} = \sum_{n=1}^\infty (-1)^n k_n^\alpha \abs{u_n}^2,
\end{equation}
for
\begin{equation}\label{eq_alpha}
\alpha = \log_\lambda \babs{\frac{c}{a}}.
\end{equation}
The $2$D parameters regime corresponds to $0 < \frac{a}{c} < 1$.
In that case the second conserved quadratic quantity $\mathrm{W}$
is identified with the enstrophy, and we can rewrite the
expression (\ref{eq_helicity}) in the form
\begin{equation} \label{eq_helicity2_new}
\mathrm{W} = \sum_{n=1}^\infty k_n^\alpha \abs{u_n}^2,
\end{equation}
where $\alpha$ is also defined by the equation (\ref{eq_alpha}).

The well-known question of global well-posedness of the
$3$D Navier-Stokes equations is a major open problem. In this work
we study analytically the shell model (\ref{eq_sabra}).
Specifically, we show global regularity of weak and strong
solutions of (\ref{eq_sabra}) and smooth dependence on the
initial data. This is a basic step for studying the long-time
behavior of this system and for establishing the existence of a
finite dimensional global attractor.

Our next result concerns the rate of the decay of the spectrum (i.
e. $\abs{u_n}$) as function of the wavenumber $k_n$. Having shown
the existence of strong solutions, we prove that $\abs{u_n}$ decays
as a polynomial in $k_n$ for $n$ large enough, depending on the
physical parameters of the model. However, if we assume that the
forcing $f_n$ decays exponentially with $k_n$, which is a reasonable
assumption, because in most numerical simulations the forcing is
concentrated in only finitely many shells, then we can show more. In
fact, following the arguments presented in \cite{FT98} and
\cite{FT89_Gevrey} we show that in this case the solution of
(\ref{eq_sabra}) belongs to a certain Gevrey class of regularity,
namely $\abs{u_n}$ also decays exponentially in $k_n$ -- an evidence
of the existence of a dissipation range associated with this system,
which is an observed feature of turbulent flows (see also
\cite{DT95_SpectrumDecay} for similar results concerning $3$D
turbulence). Similar results on the existence of the exponentially
decaying dissipation range in the GOY shell models of turbulence
were obtained previously in \cite{SchKaLo95} by the asymptotic
matching between the non-linear and viscous terms. Our results set a
mathematically rigorous ground for that claim. Moreover we show that
the exponentially decaying dissipation range exists for all values
of the parameters of the model, in particular also for $c=0$,
improving the result of \cite{SchKaLo95}.

Classical theories of turbulence assert that the turbulent flows
governed by the Navier-Stokes equations have finite number of
degrees of freedom (see, e.g., \cite{Fr95},
\cite{LL77_Hydrodynamics}). Arguing in the same vein one can state
that the sabra shell model with non-zero viscosity has finitely many
degrees of freedom. Such a physical statement can be interpreted
mathematically in more than one way: as finite dimensionality of the
global attractor, as existence of finitely many determining modes,
and as existence of finite dimensional inertial manifold.

Our first results concern the long time behavior of the solutions of
the system (\ref{eq_sabra}). We show that the global attractor of
the sabra model has a finite fractal and Hausdorff dimensions.
Moreover, we show that the number of the determining modes of the
sabra shell model equations is also finite. Remarkably, our upper
bounds for the number of the determining modes and the fractal and
Hausdorff dimensions of the global attractor of the sabra shell
model equation coincide (see Remark~\ref{rem_det_modes}). This is to
be contrasted with the present knowledge about Navier-Stokes
equations, where the two bounds are qualitatively different. In the
forthcoming paper we will investigate the lower bounds on the
Hausdorff dimension of the global attractor.

Finally, and again in contrast with the present knowledge for the
Navier-Stokes equations, we show that the sabra shell model equation
has an inertial manifold $\mathcal{M}$. Inertial manifolds are
finite dimensional Lipschitz globally invariant manifolds that
attract all bounded sets in the phase space at an exponential rate
and, in particular, contain the global attractor. We show that
$\mathcal{M} = graph(\Phi)$, where $\Phi$ is a $C^1$ function which
slaves the components $u_n$, for all $n\ge N$, as a function of $\{
u_k\}_{k=1}^N$, for $N$ large enough depending on the physical
parameters of the problem, i.e. $\nu$, $f$, $\lambda$, $a$, $b$, and
$c$. The reduction of the system (\ref{eq_sabra}) to the manifold
$\mathcal{M}$ yields a finite dimensional system of ordinary
differential equations. This is the ultimate and  best notion of
system reduction that one could hope for. In other words, an
inertial manifold is an exact rule for parameterizing the large
modes (infinite many of them) in terms of the low ones (finitely
many of them). The concept of inertial manifold for nonlinear
evolution equations was first introduced in \cite{FSTe88} (see also
\cite{ConstFoNiTe88_Inertial}, \cite{ConstFoNiTe89_Inertial2},
\cite{FST89}, \cite{SellYou92_Inertial}, \cite{Te88}).

\section{Preliminaries and Functional Setting}

Following the classical treatment of the NSE, and in order to simplify
the notation we are going to write the system (\ref{eq_sabra}) in
the following functional form
\begin{subequations}
\label{eq_abstract_model}
\begin{gather}
\pdt{u} + \nu A u + B(u, u) = f \label{eq_shell_model} \\
u(0) = u^{in}, \label{eq_initial_cond}
\end{gather}
\end{subequations}
in a Hilbert space $H$. The linear operator $A$ as well as the
bilinear operator $B$ will be defined below. In our case, the
space $H$ will be the sequences space $\ell^2$ over the field of
complex numbers $\C$. For every $u, v\in H$, the scalar product
$(\cdot, \cdot)$ and the corresponding norm $\abs{\cdot}$ are
defined as
\[
(u, v) = \sum_{n=1}^\infty u_n v_n^*, \;\;\; \abs{u} = \bigg(
\sum_{n=1}^\infty \abs{u_n}^2 \bigg)^{1/2}.
\]
We denote by $\{\phi_j\}_{j=1}^\infty$ the standard canonical
orthonormal basis of $H$, i.e. all the entries of $\phi_j$ are
zero except at the place $j$ it is equal to $1$.

The linear operator $A : D(A) \to H$ is a positive definite,
diagonal operator defined through its action on the elements of
the canonical basis of $H$ by
\[
A \phi_j = k_j^2 \phi_j,
\]
where the eigenvalues $k_j^2$ satisfy the equation (\ref{eq_freq}).
$D(A)$ - the domain of $A$ is a dense subset of $H$. Moreover, it is
a Hilbert space, when equipped with the graph norm
\[
\norm{u}{D(A)} = \abs{Au}, \;\;\; \forall u\in D(A).
\]
Using the fact that $A$ is a positive definite operator, we can
define the powers $A^s$ of $A$ for every $s\in \R$
\[
\forall u = (u_1, u_2, u_3, \dots), \;\;\; A^s u = (k_1^{2s} u_1,
k_2^{2s} u_2, k_3^{2s} u_3, \dots).
\]

Furthermore, we define the spaces
\[
V_s := D(A^{s/2}) = \{ u = (u_1, u_2, u_3, \dots) \;:\;
\sum_{j=1}^\infty k_j^{2s} \abs{u_j}^2 < \infty \},
\]
which are Hilbert spaces equipped with the scalar product
\[
(u, v)_{s} = (A^{s/2} u, A^{s/2} v), \;\;\; \forall u, v\in
D(A^{s/2}),
\]
and the norm $\abs{u}_s^2 = (u, u)_s$, for every $u\in
D(A^{s/2})$. Clearly
\[
V_{s} \subseteq V_0 = H \subseteq V_{-s}, \;\;\; \forall s > 0.
\]

The case of $s = 1$ is of a special interest for us. We denote $V
= D(A^{1/2})$ a Hilbert space equipped with a scalar product
\[
((u, v)) = (A^{1/2}u, A^{1/2}v), \;\;\; \forall u, v\in
D(A^{1/2}).
\]
We consider also $V' = D(A^{-1/2})$ - the dual space of $V$.
We denote by $\ang{\cdot, \cdot}$ the action of the functionals
from $V'$ on elements of $V$. Note that the following inclusion
holds
\[
V \subset H \equiv H' \subset V',
\]
hence the $H$ scalar product of $f\in H$ and $u\in V$ is the same
as the action of $f$ on $u$ as a functional in the duality between
$V'$ and $V$
\[
\ang{f, u} = (f, u), \;\;\; \forall f\in H, \; \forall u\in V.
\]
Observe also that for every $u\in D(A)$ and every $v\in V$ one has
\[
((u, v)) = (Au, v) = \ang{Au, v}.
\]
Since $D(A)$ is dense in $V$ one can extend the definition of the
operator $A : V \to V'$ in such a way that
\[
\ang{Au, v} = ((u, v)), \;\;\; \forall u,v \in V.
\]
In particular, it follows that
\begin{equation} \label{eq_avdual_norm}
\nm{Au}_{V'} = \nm{u}, \;\;\; \forall u\in V.
\end{equation}

Before proceeding and defining the bilinear term $B$, let us
introduce the sequence analogue of Sobolev functional spaces.

\begin{definition} \label{def_sobolev}
For $1 \le p \le \infty$ and $m\in \R$ we define sequence spaces
\[
w^{m, p} := \{ u = (u_1, u_2, \dots)\in \;:\; \nm{A^{m/2} u}_p =
\bigg( \sum_{n=1}^\infty (k_n^m \abs{u_n})^p \bigg)^{1/p} < \infty
\},
\]
for $1 \le p < \infty$, and
\[
w^{m, \infty} := \{ u = (u_1, u_2, \dots)\in \;:\; \nm{A^{m/2}
u}_\infty = \sup_{1 \le n\le \infty} (k_n^m \abs{u_n}) < \infty
\}.
\]
For $u\in w^{m, p}$ we define it's norm
\[
\nm{u}_{w^{m, p}} = \nm{A^{m/2} u}_p,
\]
where $\nm{\cdot}_p$ is the usual norm in the $\ell^p$ sequence
space. The special case of $p = 2$ and $m\ge 0$ corresponds to the
sequence analogue of the classical Sobolev space, which we denote
by
\[
h^m = w^{m, 2}.
\]
Those spaces are Hilbert with respect to the norm defined above
and its corresponding inner product.
\end{definition}

The above definition immediately implies that $h^{1} = V$. It is
also worth noting that $V \subset w^{1, \infty}$ and the inclusion
map is continuous (but not compact) because
\[
\nm{u}_{w^{1, \infty}} = \nm{A^{1/2} u}_\infty \le \nm{A^{1/2}
u}_2 = \nm{u}.
\]

The bilinear operator $B(u, v)$ will be defined formally in the
following way. Let $u, v\in H$ be of the form $u =
\sum_{n=1}^\infty u_n \phi_n$ and $v = \sum_{n=1}^\infty v_n
\phi_n$. Then
\begin{align}
B(u, v) = - i \sum_{n=1}^\infty \bigg( a k_{n+1} v_{n+2} u_{n+1}^*
& + b k_{n} v_{n+1} u_{n-1}^* + \notag \\ & + a k_{n-1} u_{n-1}
v_{n-2} + b k_{n-1} v_{n-1} u_{n-2} \bigg) \phi_n,
\end{align}
where here again $u_0 = u_{-1} = v_0 = v_{-1} = 0$. It is easy to
see that our definition of $B(u, v)$, together with the energy
conservation condition
\begin{equation*}
a + b + c = 0
\end{equation*}
implies that
\[
B(u, u) = - i \sum_{n=1}^\infty \bigg( a k_{n+1} u_{n+2} u_{n+1}^*
+ b k_{n} u_{n+1} u_{n-1}^* - c k_{n-1} u_{n-1} u_{n-2} \bigg)
\phi_n,
\]
which is consistent with (\ref{eq_sabra}). In the sequel we will
show that indeed our definition of $B(u, v)$ makes sense as an
element of $H$, whenever $u\in H$ and $v\in V$ or $u\in V$ and
$v\in H$. For $u,v \in H$ we will also show that $B(u, v)$ makes
sense as an element of $V'$.

\begin{proposition} \hfil \label{prop_non_linear}
\begin{enumerate}
    \item \label{item_vh} $B : H \times V \to H$ and $B : V \times H \to H$ are
        bounded, bilinear operators. Specifically,
        \begin{equation} \label{eq_vh}
        \abs{B(u, v)} \le C_1 \abs{u} \nm{v},
        \;\;\; \forall u\in H, v\in V,
        \end{equation}
        and
        \begin{equation} \label{eq_nl_6}
        \abs{B(u, v)} \le C_2 \nm{u} \abs{v},
        \;\;\; \forall u\in V, v\in H,
        \end{equation}
where and
\[
C_1 = (\abs{a} (\lambda^{-1} + \lambda) + \abs{b} (\lambda^{-1} +
1)),
\]
\[
C_2 = (2 \abs{a} + 2 \lambda \abs{b}).
\]

    \item \label{item_hh} $B : H \times H \to V'$ is a bounded bilinear operator
and
        \begin{equation} \label{eq_nl_1}
        \nm{B(u, v)}_{V'} \le C_1 \abs{u} \abs{v}, \;\;\; \forall u, v\in H.
        \end{equation}

    \item \label{item_hda} $B : H\times D(A) \to V$ is a bounded bilinear
operator
and for every $u\in H$ and $v\in D(A)$
        \begin{equation} \label{eq_nl_3}
        \nm{B(u, v)} \le C_3 \abs{u} \abs{A v},
        \end{equation}
where
\[
C_3 = (\abs{a} (\lambda^{3} + \lambda^{-3}) + \abs{b} (\lambda +
\lambda^{-2})).
\]

    \item \label{item_0} For every $u\in H$, $v\in V$
        \begin{equation} \label{eq_nl_2}
        Re (B(u, v), v) = 0,
        \end{equation}

\end{enumerate}
\end{proposition}

\begin{proof}
To prove the first statement, let $u\in H$ and $v\in V$. $H$ is a
Hilbert space, therefore,
\[
\abs{B(u, v)} = \sup_{w\in H, \abs{w} \le 1} \abs{(B(u, v), w)}.
\]
Using the form of the eigenvalues $k_n$ given in (\ref{eq_freq}) and
applying the Cauchy-Schwartz inequality we get
\begin{align*}
\abs{B(u, v)} & = \sup_{w\in H, \abs{w} \le 1} \babs{
\sum_{n=1}^\infty a k_{n+1} v_{n+2} u_{n+1}^* w_n^* + b k_{n}
v_{n+1} w_n^* u_{n-1}^* +
\\
& + a k_{n-1} w_n^* u_{n-1} v_{n-2} + b k_{n-1} w_n^* v_{n-1}
u_{n-2}} \le
\\
& \le (\abs{a} (\lambda^{-1} + \lambda) + \abs{b} (\lambda^{-1} +
1)) \sup_{w\in H, \abs{w} \le 1} \abs{w} \nm{v}_{w^{1, \infty}}
\abs{u} \le C_1 \abs{u} \nm{v},
\end{align*}
where the last inequality follows from the fact that $\nm{v}_{w^{1,
\infty}} \le \nm{v}$. The case of $u\in V$ and $v\in H$ is proved in
the similar way.

As for (\ref{item_hh}), let $u, v\in H$ and $w\in V$. Then
\begin{align}
\abs{\ang{B(u, v), w}} & = \babs{\sum_{n=1}^\infty B_n(u, v) w_n^*}
\le \notag
\\ & \le \babs{ \sum_{n=1}^\infty a k_{n+1} v_{n+2} u_{n+1}^* w_n^*
+ b k_{n} v_{n+1} w_n^* u_{n-1}^* + \notag \\ & + a k_{n-1} w_n^*
u_{n-1} v_{n-2} + b k_{n-1} w_n^* v_{n-1} u_{n-2}} \le \notag \\ &
\le (\abs{a} (\lambda^{-1} + \lambda) + \abs{b} (\lambda^{-1} + 1))
\nm{w}_{w^{1, \infty}} \abs{u} \abs{v} \le C_1 \nm{w} \abs{u}
\abs{v}.
\end{align}
Similar calculations can be applied to prove the statement
(\ref{item_hda}).

The statement (\ref{item_0}) follows directly from the energy
conservation condition (\ref{eq_energy_cons_assumption}).

\end{proof}

Let us fix $m \in \N$ and consider $H_m = span \{\phi_1, \dots, \phi_m\}$
and let $P_m$ be an $\ell_{2}$ orthogonal projection from $\ell^2$
onto $H_m$. If $u\in H$, then
\[
P_m u = u^m = (u_1^m, u_2^m, \dots, u_m^m, 0, 0, \dots),
\]
where $u^m_n = (u, \phi_n)$. In order to simplify the notation we
will write $u_n$ instead of $u_n^m$, whenever it will not cause
any confusion. We would like to obtain the same estimates as above
in Proposition~\ref{prop_non_linear} for the truncated bilinear
term $P_m B(u^m, v^m)$. Setting $u_j = u^m_j = 0$ for all $j = 0,
-1$ and $j > m$ we get
\begin{align*}
P_m B(u^m, v^m) = - i & \sum_{n = 2}^{m-1} \bigg( a k_{n+1}
v_{n+2} u_{n+1}^* \phi_{n-1} + b k_{n} v_{n+1} u_{n-1}^* \phi_{n}
+ \notag
\\ & + a k_{n-1} u_{n-1} v_{n-2} \phi_{n+1} + b k_{n-1} v_{n-1}
u_{n-2} \phi_{n+1} \bigg).
\end{align*}

We observe that similar estimates to (\ref{eq_nl_1}),
(\ref{eq_nl_vdnorm}) and (\ref{eq_nl_2}) hold for the truncated
bilinear operator $P_m B(u^m, v^m)$. Namely, $\forall u^m, v^m\in
H_m$
\begin{equation} \label{eq_nl_vdnorm}
\norm{P_m B(u^m, v^m)}{V'} \le C_1 \abs{u^m} \abs{v^m}.
\end{equation}
and $\forall u^m, v^m\in H$
\begin{equation} \label{eq_nl_findim_2}
Re (P_m B(u^m, v^m), v^m) = 0.
\end{equation}

\section{Existence of Weak Solutions}

In this section we prove the existence of weak solutions to the
system (\ref{eq_abstract_model}). Our notion of weak solutions is
similar to that of the Navier-Stokes equation (see, for example,
\cite{CF88}, \cite{Te84}).

\begin{theorem} \label{thm_weak_solution}
Let $f\in \l{}{2}([0, T), V')$ and $u^{in} \in H$. There exists
\begin{equation} \label{eq_u_bounds_weak}
u\in \l{}{\infty}([0, T], H) \cap \l{}{2}([0, T], V),
\end{equation}
with
\begin{equation} \label{eq_dudt_bounds_weak}
\pdt{u} \in \l{}{2}([0, T], V'),
\end{equation}
satisfying the weak formulation of the equation
(\ref{eq_abstract_model}), namely
\begin{subequations}
\label{eq_abstract_model_weak}
\begin{gather}
\ang{\pdt{u}, v} + \nu \ang{A u, v} + \ang{B(u, u), v} = \ang{f, v}
\label{eq_shell_model_weak} \\ u(x, 0) = u^{in}(x),
\label{eq_initial_cond_weak}
\end{gather}
\end{subequations}
for every $v\in V$.

The weak solution $u$ satisfies
\begin{equation} \label{eq_u_bounds_weak_cont}
u\in C([0, T], H).
\end{equation}
\end{theorem}

\begin{proof}
The proof follows standard techniques for equations of Navier-Stokes
type. See, for example, (\cite{CF88}, \cite{Te84} or \cite{Te88}).
We will use the Galerkin procedure to prove the global existence and
to establish the necessary \emph{a priori} estimates.

The Galerkin approximating system of order $m$ for equation
(\ref{eq_abstract_model}) is an $m$ - dimensional system of ordinary
differential equations
\begin{subequations}
\label{eq_abstract_model_galerkin}
\begin{gather}
\frac{du^m}{dt} + \nu A u^m + P_m B(u^m, u^m) = P_m f
\label{eq_galerkin} \\ u^m(0) = P_m u^{in}. \label{eq_galerkin_ic}
\end{gather}
\end{subequations}
We observe that the nonlinear term is quadratic in $u^m$, thus by
the theory of ordinary differential equations, the system
(\ref{eq_abstract_model_galerkin}) possesses a unique solution for
a short interval of time. Denote by $[0, T_m)$ the maximal
interval of existence for positive time. We will show later that
in fact, $T_m = T$.

Let us fix $m$ and the time interval $[0, T]$. On $[0, T_m)$ we take
the inner product $(\cdot, \cdot)_H$ of the equation
(\ref{eq_galerkin}) with $u^m$, considering the real part and using
(\ref{eq_nl_findim_2}) we get
\[
\frac{1}{2} \ddt \abs{u^m}^2 + \nu \nm{u^m}^2 = Re (P_m f, u^m).
\]
The right hand side could be estimated using the Cauchy-Schwartz and
Young inequalities
\[
\abs{(P_m f, u^m)} \le \nm{f}_{V'} \nm{u^m} \le \frac{1}{2\nu}
\nm{f}_{V'}^2 + \frac{\nu}{2} \nm{u^m}^2.
\]
Plugging it into the equation we get
\begin{equation}\label{eq_ineq}
\frac{d}{dt} \abs{u^m}^2 + \nu \nm{u^m}^2 \le \frac{1}{\nu}
\nm{f}_{V'}^2.
\end{equation}

Integrating the last equation from $0$ to $s$, where $0 < s \le
T_m$, we conclude that for every such $s$,
\[
\abs{u^m(s)}^2 \le \abs{P_m u^{in}}^2 + \frac{1}{\nu} \int_0^T
\nm{f(t)}_{V'} dt.
\]
Thus, $\limsup_{s\to T_m^-} \abs{u_m(s)}^2 < \infty$. Therefore,
the maximal interval of existence of solution of the system
(\ref{eq_galerkin}) for every $m$ is $[0, T)$ - an interval, where
the terms of the equation make sense, and in particular, $f$. If
$f\in \l{}{2}([0, \infty), V')$, then we can take $T > 0$
arbitrary large and show the existence of weak solutions on $[0,
T)$. Now, we observe that
\[
\nm{u^m}_{\l{}{\infty}([0, T), H)}^2 \le \abs{P_m u^{in}}^2 +
\frac{1}{\nu} \int_0^T \nm{f}_{V'} dt,
\]
and hence
\begin{equation}\label{eq_conclusion1}
\text{The sequence } u^m \text{ lies in the bounded set of }
\l{}{\infty}([0, T], H).
\end{equation}

Next, integrating (\ref{eq_ineq}) over the interval $[0, T]$ we
get
\[
\abs{u^m(T)}^2 + \nu \int_0^T \nm{u^m(t)}^2 dt \le \abs{u^m(0)}^2
+ \frac{1}{\nu} \int_0^T \nm{f(t)}_{V'} dt,
\]
and using the fact that $\abs{u^m(0)} = \abs{P_m u^{in}} \le
\abs{u^{in}}$ we conclude
\begin{equation} \label{eq_conclusion2}
\text{The sequence } u^m \text{ lies in the bounded set of }
\l{}{2}([0, T], V).
\end{equation}

Finally, in order to apply Aubin's Compactness Lemma (see, e.g.,
\cite{Au63}, \cite{CF88}) we need to estimate the norm
$\norm{\pdt{u_m}}{V'}$. Consider the equation
\[
\frac{du^m}{dt} = - \nu A u^m - P_m B(u^m, u^m) + P_m f
\]
Then, (\ref{eq_nl_vdnorm}) and (\ref{eq_conclusion1}) imply that
$\norm{P_m B(u^m, u^m)}{V'}$ is uniformly bounded for every $0 \le
t \le T$, hence
\[
P_m B(u^m, u^m) \in \l{}{p}([0, T], V'),
\]
for every $p > 1$ and $T < \infty$. Moreover, from the fact that
$\norm{P_m f}{V'} \le \norm{f}{V'}$ and the condition on $f$ it
follows that $P_m f\in \l{}{2}([0, T], V')$.

The relations (\ref{eq_avdual_norm}) and (\ref{eq_conclusion2})
imply that the sequence $\{A u^m\}$ is uniformly bounded in the
space $\l{}{2}([0, T], V')$. Therefore, using the fact that $f\in
\l{}{2}([0, T], V')$ we conclude,
\begin{equation} \label{eq_conclusion3}
\text{The sequence } \frac{du^m}{dt} \text{ lies in the bounded set
of } \l{}{2}([0, T], V').
\end{equation}

At last, we are able to apply Aubin's Compactness Lemma. According to
(\ref{eq_conclusion1}), (\ref{eq_conclusion2}) and
(\ref{eq_conclusion3}), there exists a subsequence $u^{m'}$, and
$u\in H$ satisfying
\begin{subequations}
\label{eq_converge}
\begin{gather}
u^{m'} \longrightarrow u \;\; \text{ weakly in } \;\; \l{}{2}([0,
T], V) \label{eq_converge_weakly} \\ u^{m'} \longrightarrow u \;\;
\text{ strongly in } \;\; \l{}{2}([0, T], H)
\label{eq_converge_strongly}
\end{gather}
\end{subequations}

We are left to show that the solution $u$ that we found,
satisfies the weak formulation of the equation
(\ref{eq_abstract_model}) together with the initial condition. Let
$v\in V$ be an arbitrary function. Take the scalar product of
(\ref{eq_galerkin}) with $v$ and integrate for some $0 \le t_0 < t
\le T$ to get
\begin{align}
(u^{m'}(t), v) - (u^{m'}(t_0), v) & + \nu \int_{t_0}^t
((u^{m'}(\tau), v)) d \tau + \notag \\ & + \int_{t_0}^t (P_{m'}
B(u^{m'}, u^{m'}), v) d \tau = \int_{t_0}^t \ang{P_{m'} f, v} d
\tau. \label{eq_limit}
\end{align}

The conclusion (\ref{eq_converge}) implies that there exists a set
$E\subset [0, T]$ of Lebesgue measure $0$ and a subsequence
$u^{m''}$ of $u^{m'}$ such that for every $t\in [0, T]\setminus
E$, $u^{m''}(t)$ converges to $u(t)$ weakly in $V$ and strongly in
$H$. As for the third term
\[
\lim_{m' \to \infty} \int_{t_0}^t ((u^{m'}(\tau), v)) d \tau
\longrightarrow \int_{t_0}^t ((u(\tau), v)) d \tau,
\]
because of (\ref{eq_converge_weakly}). The same holds for the
forcing term. We are left to evaluate the nonlinear part. It is
enough to check that
\begin{equation}
\int_{t_0}^t \abs{(P_{m'} B(u^{m'}, u^{m'}), v) - (B(u, u), v)} d
\tau \longrightarrow 0, \;\;\; \text{as} \; m'\to \infty.
\label{eq_nl_converge}
\end{equation}

Let us define $Q = Q_{m'} = I - P_{m'}$. Then, by definition of
$B(u, u)$ and arguments similar to (\ref{eq_nl_1}) we have
\begin{align*}
\abs{(P_{m'} & B(u^{m'}, u^{m'}), v) - B(u, u), v)} = \\ & =
\babs{\sum_{n = m}^\infty \bigg( a k_{n+1} v_{n+2} u_{n+1}^*
v_n^* + b k_{n} v_{n+1} v_n^* u_{n-1}^* + a k_{n-1} v_n^* u_{n-1} v_{n-2} + \\
& + b k_{n-1} v_n^* v_{n-1} u_{n-2} \bigg)} \le C \nm{Q v} \abs{Q
u}^2 \le C \nm{v} \abs{u - P_{m'} u}^2.
\end{align*}
We see that (\ref{eq_nl_converge}) holds, because of
(\ref{eq_converge_strongly}). Passing to the limit in
(\ref{eq_limit}) we finally conclude that $u$ is weakly continuous
in $H$ and that it satisfies the weak form of the equation, namely
\begin{align}
(u(t), v) - (u(t_0), v) & + \nu \int_{t_0}^t ((u(\tau), v)) d \tau +
\notag \\ & + \int_{t_0}^t (B(u, u), v) d \tau = \int_{t_0}^t
\ang{f, v} d \tau. \notag
\end{align}

In order to finish the proof we need to show that $u$ satisfies
(\ref{eq_u_bounds_weak_cont}). First, recall that $u\in
\l{}{2}([0, T], V)$ and $\pdt{u} \in \l{}{2}([0, T], V')$.
Therefore, it follows that $\abs{u(t)}^2$ is continuous. Moreover,
$u\in C_w([0, T], H)$ -- the space of weakly continuous functions
with values in $H$. Now, let us fix $t\in [0, T]$ and consider the
sequence $\{t_n\}\subset [0, T]$ converging to $t$. We know that
$\abs{u(t_n)}$ converges to $\abs{u(t)}$ and $u(t_n)$ converges
weakly to $u(t)$. Then, it follows that $u(t_n)$ converges
strongly to $u(t)$, and we finally conclude that
\[
u\in C([0, T], H).
\]

\end{proof}

\section{Uniqueness of Weak Solutions}

Next we show that the weak solutions depend continuously on the
initial data and in particular the solutions are unique.

\begin{theorem} \label{thm_cont_dependence}
Let $u(t), v(t)$ be two different solutions to the equation
(\ref{eq_abstract_model}) on the time interval $[0, T]$ with the
corresponding initial conditions $u^{in}$ and $v^{in}$ in $H$.
Then, for every $t \in [0, T]$ we have
\begin{equation} \label{eq_dependence}
\abs{u(t) - v(t)} \le e^{K} \abs{u^{in} - v^{in}},
\end{equation}
where
\begin{equation} \label{eq_k_constant}
K = C_1 \int_0^t \nm{u(s)} ds.
\end{equation}
\end{theorem}

\begin{proof}
\noindent Consider $u, v$ -- two solutions of the equation
(\ref{eq_shell_model}) satisfying the initial conditions $u(0) =
u^{in}$ and $v(0) = v^{in}$. The difference $w = u - v$ satisfies
the equation
\begin{equation*}
\frac{d w}{dt} + \nu Aw + B(u, w) + B(w, u) - B(w, w) = 0,
\end{equation*}
with the initial condition $w(0) = w^{in} = u^{in} - v^{in}$.

According to Theorem~\ref{thm_weak_solution}, $\pdt{w} \in
\l{}{2}([0, T), V')$ and $w \in \l{}{2}([0, T), V)$. Hence, the
particular case of the general theorem of interpolation due to
Lions and Magenes~\cite{LM72} (see also~\cite{Te84}, Chap. III,
Lemma $1.2$), implies that $\ang{\pdt{w}, w}_{V'} = \half \ddt
\abs{w}^2$. Taking the inner product $<\cdot, w>$ and using
(\ref{eq_nl_1}) and (\ref{eq_nl_2}) we conclude that
\begin{equation*}
\frac{1}{2} \ddt \abs{w}^2 \le \frac{1}{2} \ddt \abs{w}^2 + \nu
\nm{w}^2 = - Re <B(w, u), w> \le C_1 \nm{u} \abs{w}^2.
\end{equation*}
Gronwall's inequality now implies
\begin{equation}
\abs{w(t)}^2 \le \abs{w^{in}}^2 \exp \bigg( C_1 \int_0^t \nm{u(s)}
ds \bigg). \label{eq_uniqueness}
\end{equation}
Since $w^{in} = u^{in} - v^{in}$ and $u\in \l{}{2}([0, T), V)$ the
equation (\ref{eq_dependence}) follows. In particular, in the case
that $u^{in} = v^{in}$, it follows that $u(t) = v(t)$ for all $t \in
[0, T]$.

\end{proof}

\section{Strong Solutions}

In the analogy to the Navier-Stokes equations we would like to
show that the shell model equation (\ref{eq_abstract_model})
possesses even more regular solutions under an appropriate
assumptions.

Let $u^m$ be the solution of the Galerkin system
(\ref{eq_abstract_model_galerkin}). Taking the scalar product of the
equation (\ref{eq_galerkin}) with $A u^m$, and using Young and
Cauchy-Schwartz inequalities we obtain
\begin{align*}
\half \ddt \nm{u^m}^2 + \nu \abs{Au^m}^2 & \le \abs{(P_m B(u^m,
u^m), A u^m)} + \abs{(f^m, A u^m)} \le \\ & \le C_1 \nm{u^m}
\abs{u^m} \abs{A u^m} + \frac{1}{\nu} \abs{f}^2 + \frac{\nu}{4}
\abs{A u^m}^2 \le \\ & \le \frac{C_1^2}{\nu} \nm{u^m}^2 \abs{u^m}^2
+ \frac{\nu}{4} \abs{A u^m}^2 + \frac{1}{\nu} \abs{f}^2 +
\frac{\nu}{4} \abs{A u^m}^2.
\end{align*}
Finally, we get
\begin{equation*}
\ddt \nm{u^m}^2 + \nu \abs{Au^m}^2 \le \frac{2 C_1^2}{\nu}
\nm{u^m}^2 \abs{u^m}^2 + \frac{2}{\nu} \abs{f}^2.
\end{equation*}

By omitting for the moment the term $\nu \abs{Au^m}^2$ from the
left hand side of the equation, we can multiply it by the
integrating factor $\exp(- \frac{2 C_1^2}{\nu} \int_{t_0}^t
\abs{u^m(s)}^2 ds)$ for some $0 \le t_0 \le t$ to get
\begin{equation}
\ddt \bigg( \nm{u^m}^2 \exp(- \frac{2 C_1^2}{\nu} \int_{t_0}^t
\abs{u^m(s)}^2 ds) \bigg) \le \frac{2}{\nu} \abs{f}^2 \exp(- \frac{2
C_1^2}{\nu} \int_{t_0}^t \abs{u^m(s)}^2 ds) \le \frac{2}{\nu}
\abs{f}^2. \label{eq_strong_1}
\end{equation}

Assuming that the forcing $f$ is essentially bounded in the norm
of $H$ for all $0 \le t \le T$, namely that $f\in \l{}{\infty}([0,
T], H)$ and denoting $\abs{f}_{\infty} = \nm{f}_{\l{}{\infty}([0,
T], H)}$, we obtain by integrating equation (\ref{eq_strong_1})
\begin{align} \label{eq_strong_main}
\nm{u^m(t)}^2 \le e^{( \frac{2 C_1^2}{\nu} \int_{t_0}^t
\abs{u^m(s)}^2 ds )} \bigg( \nm{u^m(t_0)}^2 + \frac{2}{\nu}
\abs{f}_\infty^2 (t - t_0) \bigg).
\end{align}

In order to obtain a uniform bound on $\nm{u^m(t)}^2$ for all $0
\le t \le T$ and for all $m$ we need to estimate both the exponent
and $\nm{u^m(t_0)}^2$. The last quantity is of course bounded for
$t_0 = 0$ if we assume that $P_m u^{in}\in V$ for all $m$, however
we want to put a less strict assumptions. Let us first refine
the estimates we made in the proof of the existence of weak
solutions. Take the scalar product of equation (\ref{eq_galerkin})
with $u^m$ to get
\[
\half \ddt \abs{u^m}^2 + \nu \nm{u^m}^2 = (f^m, u^m) \le \frac{1}{2
k_1^2 \nu} \abs{f}^2 + \frac{k_1^2 \nu}{2} \abs{u^m}^2,
\]
and using the inequality $k_1^2 \abs{u^m}^2 \le \nm{u^m}^2$ we get
\begin{equation}
\ddt \abs{u^m}^2 + \nu \nm{u^m}^2 \le \frac{1}{k_1^2 \nu} \abs{f}^2.
\label{eq_strong_22}
\end{equation}
Integrating from $0$ to $t$, and once again using the fact that
$f\in \l{}{\infty}([0, T], H)$, we get
\begin{equation} \label{eq_strong_2}
\nu \int_0^t \nm{u^m(s)}^2 ds \le \abs{u^m(0)}^2 + \frac{1}{k_1^2
\nu} \abs{f}_\infty^2 t.
\end{equation}
On the other hand, Gronwall's inequality, applied to
(\ref{eq_strong_22}) gives us
\begin{equation*}
\abs{u^m(t)}^2 \le \abs{u^m(0)}^2 e^{-\nu k_1^2 t} + \int_0^t
e^{-\nu k_1^2 (t - s)} \frac{\abs{f}^2}{k_1^2 \nu} ds,
\end{equation*}
and implies
\begin{equation} \label{eq_strong_4}
\abs{u^m(t)}^2 \le \abs{u^m(0)}^2 e^{- k_1^2 \nu t} +
\frac{\abs{f}_\infty^2}{k_1^4 \nu^2} (1 - e^{- k_1^2 \nu t}),
\end{equation}
where the last expression yields for every $t > 0$
\begin{equation} \label{eq_strong_3}
\abs{u^m(t)}^2 \le \abs{u^m(0)}^2 + \frac{\abs{f}_\infty^2}{k_1^4
\nu^2}.
\end{equation}

Using the equation (\ref{eq_strong_3}) we immediately can evaluate
the exponent in (\ref{eq_strong_main}) by
\begin{equation} \label{eq_strong_4_new}
e^{( \frac{2 C_1^2}{\nu} \int_{t_0}^t \abs{u^m(s)}^2 ds )} \le e^{
\frac{2 C_1^2}{\nu} (t - t_0) (\abs{u^m(0)}^2 + \frac{1}{k_1^4
\nu^2} \abs{f}_\infty^2 )}.
\end{equation}

The last step is to get a uniform bound for the $\nm{u^m(t_0)}^2$.
Let us integrate equation (\ref{eq_strong_22}) from $t$ to $t +
\tau$. The result, after applying (\ref{eq_strong_3}) is
\begin{align*}
\nu \int_t^{t + \tau} \nm{u^m}^2 ds & \le \abs{u^m(t + \tau)}^2 -
\abs{u^m(t)}^2 + \frac{\tau}{k_1^2 \nu} \abs{f}_\infty^2 \le \\ &
\le \abs{u^m(0)}^2 + \frac{1}{k_1^2 \nu} \abs{f}_\infty^2 \bigg(
\frac{1}{k_1^2 \nu} + \tau \bigg)
\end{align*}
The standard application of Markov's inequality implies that on
every interval of length $\tau$ there exists a time $t_0$ satisfying
\begin{equation}
\nm{u^m(t_0)}^2 \le \frac{2}{\tau} \bigg( \frac{1}{\nu}
\abs{u^m(0)}^2 + \frac{1}{k_1^2 \nu^2} \abs{f}_\infty^2 \bigg(
\frac{1}{k_1^2 \nu} + \tau \bigg) \bigg). \label{eq_strong_5}
\end{equation}

Let us set $\tau \le \frac{1}{k_1^2 \nu}$. Then, for every $t \ge
\tau$ and an appropriate $t_0$ chosen in the interval $[t - \tau,
t]$ we get a uniform bound for all $m\in \N$ and all $t\ge \tau$, by
substituting (\ref{eq_strong_4}) and (\ref{eq_strong_5}) into
(\ref{eq_strong_main})
\begin{align} \label{eq_strong_main3}
\tau \nm{u^m(t)}^2 \le 2 e^{ \frac{2 C_1^2}{k_1^2 \nu^2}
(\abs{u^m(0)}^2 + \frac{1}{k_1^4 \nu^2} \abs{f}_\infty^2 )} \cdot
\bigg( \frac{1}{\nu} \abs{u^m(0)}^2 + \frac{3}{k_1^4 \nu^3}
\abs{f}_\infty^2 \bigg)
\end{align}

To expand the estimate for all times $0 \le t \le \frac{1}{k_1^2
\nu}$, we can apply the last inequality to the dyadic intervals of
the form $[\frac{1}{k_1^2 \nu} \frac{1}{2^{k+1}}, \frac{1}{k_1^2
\nu} \frac{1}{2^{k}}]$ with $\tau = \frac{1}{k_1^2 \nu}
\frac{1}{2^{k+1}}$. Then, if $t\in [\frac{1}{k_1^2 \nu}
\frac{1}{2^{k+1}}, \frac{1}{k_1^2 \nu} \frac{1}{2^{k}}]$, it
satisfies $\tau \le t \le 2\tau$, and for such $t$ inequality
(\ref{eq_strong_main3}) implies
\begin{align} \label{eq_strong_main3_new}
t \nm{u^m(t)}^2 \le 4 e^{ \frac{2 C_1^2}{k_1^2 \nu^2}
(\abs{u^m(0)}^2 + \frac{1}{k_1^4 \nu^2} \abs{f}_\infty^2 )} \cdot
\bigg( \frac{1}{\nu} \abs{u^m(0)}^2 + \frac{3}{k_1^4 \nu^3}
\abs{f}_\infty^2 \bigg)
\end{align}
Since the last estimate does not depend on $k$ we can conclude that
it holds for all $0 \le t \le \frac{1}{k_1^2 \nu}$.

Finally, passing to the limit in the Galerkin approximation we
obtain

\begin{theorem} \label{thm_strong}
Let $T > 0$ and $f\in \l{}{\infty}([0, T], H)$. Then if $u^{in}\in
H$, the equation (\ref{eq_abstract_model}) possesses a solution
$u(t)$ satisfying
\[
u\in \l{loc}{\infty}((0, T], V) \bigcap \l{loc}{2}((0, T], D(A))
\bigcap \l{}{\infty}([0, T], H) \bigcap \l{}{2}([0, T], V).
\]
Moreover, the following bound holds
\begin{align}
\sup_{0 < t \le \min\{\frac{1}{k_1^2 \nu}, T\}} \nu k_1^2 t
\nm{u(t)}^2 & + \sup_{\frac{1}{k_1^2 \nu} \le t \le T} \nm{u(t)}^2
\le \notag
\\ & \le 6 e^{ \frac{2 C_1^2}{k_1^2 \nu^2} (\abs{u^{in}}^2 +
\frac{1}{k_1^4 \nu^2} \abs{f}_\infty^2 )} \cdot \bigg( k_1^2
\abs{u^{in}}^2 + \frac{3}{k_1^2 \nu^2} \abs{f}_\infty^2 \bigg).
\label{eq_strong_bound}
\end{align}

If the initial condition $u^{in}\in V$, then the solution $u(t)$
satisfies
\begin{equation} \label{eq_u_strong_estimate}
u\in C([0, T], V) \bigcap \l{}{2}([0, T], D(A)),
\end{equation}
\end{theorem}

\section{Gevrey class regularity}

\subsection{Preliminaries and Classical Navier-Stokes Equation Results}

We start this section with a definition

\begin{definition}
Let us fix positive constants $\tau, p > 0$, $q \ge 0$ and define
the following class of sequences
\[
G^{p, q}_\tau = \{ u\in \ell_2 \;:\; \abs{e^{\tau A^{p/2}} A^{q/2}
u} = \sum_{n=1}^\infty e^{2 \tau k_n^p} k_n^q \abs{u_n}^2 < \infty
\}.
\]
We will say that the sequence $u\in \ell_2$ is Gevrey regular if it
belongs to the class $G^{p, q}_\tau$ for some choice of $\tau, p >
0$ and $q \ge 0$.
\end{definition}

In order to justify this definition let $u = (u_1, u_2, u_3,
\dots)$ and suppose that $u_n$, $n = 1, 2, 3, \dots$, are the
Fourier coefficients of some scalar function $g(x)$, with
corresponding wavenumbers $k_n$, defined by relation
(\ref{eq_freq}). Then, if $u\in G^{p, q}_\tau$, for some $p, q$
and $\tau$, then the function $g(x)$ is analytic. The concept of
the Gevrey class regularity for showing the analyticity of the
solutions of the two-dimensional Navier-Stokes equations, was
first introduced in \cite{FT89_Gevrey}, simplifying earlier
proofs. Later this technique was extended to the large class of
analytic nonlinear parabolic equations in \cite{FT98}.

For $q = 0$ we will write $G^p_\tau = G^{p, 0}_\tau$. Denote by
$(\cdot, \cdot)_{p, \tau}$ the scalar product in $G^p_\tau$ and
\[
\abs{u}_{p, \tau} = \nm{u}_{G^{p}_\tau}, \;\;\; \forall u\in
G^{p}_\tau.
\]
In the case of $q = 1$ we will write
\[
\nm{u}_{p, \tau} = \nm{u}_{G^{p, 1}_\tau}, \;\;\; \forall u\in G^{p,
q}_\tau,
\]
and the scalar product in $G^{p, 1}_\tau$ will be denoted by
$((\cdot, \cdot))_{p, \tau}$.

Following the tools introduced in \cite{FT89_Gevrey} and generalized in
\cite{FT98} we prove the following:

\begin{theorem} \label{thm_gevrey}
Let us assume that $u(0) = u^{in} \in V$ and the force $f\in
G^{p_1}_{\sigma_1}$ for some $\sigma_1, p_1 > 0$. Then, there
exists $T$, depending only on the initial data, such that for
every $0 < p \le \min \{p_1, \log_\lambda \frac{1 + \sqrt{5}}{2}
\}$ the following holds
\begin{enumerate}
    \item For every $t \in [0, T]$ the solution $u(t)$ of the
    equation (\ref{eq_abstract_model}) remains bounded in
    $G^{p, 1}_{\psi(t)}$, for $\psi(t) = \min \{t, \sigma_1 \}$.

    \item Furthermore, there exists $\sigma > 0$ such that for every $t > T$
    the solution $u(t)$ of the equation (\ref{eq_abstract_model})
    remains bounded in $G^{p, 1}_\sigma$.
\end{enumerate}
\end{theorem}

Before proving the theorem, we prove the following estimates for
the nonlinear term:

\begin{lemma} \label{lem_gevrey}
Let $u, v, w\in G^{p, 1}_\tau$ for some $\tau > 0$ and $p \le
\log_\lambda \frac{1 + \sqrt{5}}{2}$. Then $B(u, v)$ belongs to
$G^p_\tau$ and satisfies
\begin{equation} \label{eq_nl_gevrey}
\abs{(A^{1/2} e^{\tau A^{p/2}} B(u, v), A^{1/2} e^{\tau A^{p/2}}
w)} \le C_4 \nm{u}_{G^{p, 1}_\tau} \nm{v}_{G^{p, 1}_\tau}
\nm{w}_{G^{p, 1}_\tau},
\end{equation}
for an appropriate constant $C_4 > 0$.
\end{lemma}

\begin{proof}

Let us fix an arbitrary $p \le \log_\lambda \frac{1 +
\sqrt{5}}{2}$. Observe, that the operator $e^{\tau A^{p/2}}$ is
diagonal in the standard canonical basis $\{\phi_n
\}_{n=1}^\infty$ of $H$ with corresponding eigenvalues $e^{\tau
\widetilde{k}_n}$, $n = 1, 2, \dots$, where $\widetilde{k}_n =
\widetilde{\lambda}^n$ for some $\widetilde{\lambda} \le \frac{1 +
\sqrt{5}}{2}$. Therefore,
\begin{align}
\abs{(A^{1/2} & e^{\tau A^{p/2}} B(u, v), A^{1/2} e^{\tau A^{p/2}}
w)} \le \notag \\ & \le \sum_{n=1}^\infty e^{2 \tau \widetilde{k}_n}
\babs{ a k_{n+1} k_{n}^2 v_{n+2} u_{n+1}^* w_n^* + b k_{n}^3 v_{n+1}
w_n^* u_{n-1}^* \notag \\ & + a k_n^2 k_{n-1} w_n^* u_{n-1} v_{n-2}
+ b k_n^* k_{n-1} w_n^* v_{n-1} u_{n-2}} \le \notag \\ & \le \sup_{n
\ge 1} \abs{e^{\tau \widetilde{k_n}} k_n w_n } \sum_{n=1}^\infty
e^{\tau \widetilde{k}_n} \bigg( \babs{ a k_{n+1} k_n v_{n+2}
u_{n+1}^* } + \babs{ b k_{n}^2 v_{n+1} u_{n-1}^* } + \notag \\ & +
\babs{ a k_n k_{n-1} u_{n-1} v_{n-2} } + \babs{ b k_n k_{n-1}
u_{n-2} v_{n-1} } \bigg) \le C_4 \nm{w}_{G^{p, 1}_\tau}
\nm{u}_{G^{p, 1}_\tau} \nm{v}_{G^{p, 1}_\tau},
\end{align}
where we can choose $C_4 = (\abs{a} (\lambda^{-2} + \lambda^{2}) +
\abs{b} (1 + \lambda^{2}))$. The last inequality is the result of
the Cauchy-Schwartz inequality and the fact, that $e^{\tau
(\widetilde{\lambda}^n - \widetilde{\lambda}^{n+2} -
\widetilde{\lambda}^{n+1})}$, $e^{\tau (\widetilde{\lambda}^n -
\widetilde{\lambda}^{n+1} - \widetilde{\lambda}^{n-1})}$ and
$e^{\tau (\widetilde{\lambda}^n - \widetilde{\lambda}^{n-1} -
\widetilde{\lambda}^{n-2})}$ are less than or equal to $1$ for all
$\tau > 0$ and our specific choice of $\widetilde{\lambda}$.

\end{proof}

Now we are ready to prove the Theorem:

\begin{proof} (Of Theorem~\ref{thm_gevrey})
Let us take an arbitrary $0 < p \le \min \{p_1, \log_\lambda
\frac{1 + \sqrt{5}}{2} \}$ and denote $\varphi(t) = \min(t,
\sigma_1)$. For a fixed time $t$ take the scalar product of the
equation (\ref{eq_shell_model}) with $A u(t)$ in
$G^{p}_{\varphi(t)}$ to get
\begin{align} \label{eq_gevrey_scalar_product}
(u'(t), A u(t))_{p, \varphi(t)} & + \nu \abs{A u(t)}_{p,
\varphi(t)}^2 = \notag \\ & = (f, A u(t))_{p, \varphi(t)} -
(B(u(t), u(t)), A u(t))_{p, \varphi(t)}.
\end{align}

The left hand side of the equation could be transformed in the
following way:
\begin{align} \label{eq_gevrey_left side}
(u'(t), & A u(t))_{p, \varphi(t)} = (e^{\varphi(t) A^{p/2}} u'(t),
e^{\varphi(t) A^{p/2}} A u(t)) = \notag \\ & = (A^{1/2}
(e^{\varphi(t) A^{p/2}} u(t))' - \varphi'(t) A^{(p+1)/2}
e^{\varphi(t) A^{p/2}} u(t), A^{1/2} e^{\varphi(t) A^{p/2}} u(t))
= \notag \\ & = \half \ddt \nm{u(t)}^2_{p, \varphi(t)} -
\varphi'(t) (A^{(p+1)/2} u(t), A^{1/2} u(t))_{p, \varphi(t)} \ge
\notag \\ & \ge \half \ddt \nm{u(t)}^2_{p, \varphi(t)} - \abs{A
u(t)}_{p, \varphi(t)} \nm{u(t)}_{p, \varphi(t)} \ge \notag \\ &
\ge \half \ddt \nm{u(t)}^2_{p, \varphi(t)} - \frac{\nu}{4} \abs{A
u(t)}_{p, \varphi(t)}^2 - \frac{1}{\nu} \nm{u(t)}_{p,
\varphi(t)}^2.
\end{align}

The right hand side of the equation (\ref{eq_gevrey_scalar_product})
can be bounded from above in the usual way using
Lemma~\ref{lem_gevrey}, Cauchy-Schwartz and Young inequalities
\begin{align} \label{eq_gevrey_right_side}
\abs{(f, A u(t))_{p, \varphi(t)} & - (B(u(t), u(t)), A u(t))_{p,
\varphi(t)}} \le \notag \\ & \le \abs{f}_{p, \varphi(t)} \abs{A
u(t)}_{p, \varphi(t)} + C_4 \nm{u}_{p, \varphi(t)}^3 \le \notag \\
& \le \frac{1}{\nu} \abs{f}_{p, \varphi(t)}^2 + \frac{\nu}{4} \abs{A
u(t)}_{p, \varphi(t)}^2 + C_4 \nm{u}_{p, \varphi(t)}^3.
\end{align}

Combining (\ref{eq_gevrey_left side}) and
(\ref{eq_gevrey_right_side}), the equation
(\ref{eq_gevrey_scalar_product}) takes the form
\begin{align*}
\ddt \nm{u(t)}^2_{p, \varphi(t)} + \nu \abs{A u(t)}_{p,
\varphi(t)}^2 & \le \frac{2}{\nu} \abs{f}_{p, \varphi(t)}^2 +
\frac{2}{\nu} \nm{u(t)}_{p, \varphi(t)}^2 + C_4 \nm{u}_{p,
\varphi(t)}^3 \le \\ & \le \frac{2}{\nu} \abs{f}_{p, \varphi(t)}^2 +
C_5 + C_4 \nm{u}_{p, \varphi(t)}^3.
\end{align*}

Denoting $y(t) = 1 + \nm{u(t)}_{p, \varphi(t)}^2$ and $K =
\frac{2}{\nu} \abs{f}_{p, \varphi(t)}^2 + C_5 + C_4^{2/3}$ we get an
inequality
\[
\dot{y} \le K y^{3/2},
\]
It implies
\[
y(t) = 1 + \nm{u(t)}_{p, \varphi(t)}^2 \le 2 y(0) = 2(1 +
\nm{u^{in}}^2),
\]
for
\[
t \le T_1(\sigma_1, \nm{u^{in}}, \abs{f}_{p, \varphi(t)}) = \frac{2
- \sqrt{2}}{K} y^{-1/2}(0) = \frac{2 - \sqrt{2}}{K} (1 +
\nm{u^{in}}^2)^{-1/2}.
\]

According to Theorem~\ref{thm_strong}, the solution $u(t)$ of the
equation (\ref{eq_abstract_model}) remains bounded in $V$ for all
time if we start from $u^{in}\in V$. Moreover, from the relation
(\ref{eq_u_strong_estimate}) we conclude that there exists a
constant $M$, such that for all $t > 0$
\[
\nm{u(t)} \le M.
\]
Hence we can repeat the above arguments starting from any time $t
> 0$ and find that
\begin{equation} \label{eq_gevrey_last}
\abs{e^{\varphi(T_2) A^{p/2}} A^{1/2} u(t)} \le 2 + 2 M^2,
\end{equation}
for all $t\ge T_2 = \frac{2 - \sqrt{2}}{K} (1 + M^2)^{-1/2}$, and
the Theorem holds.

\end{proof}

\subsection{A Stronger Result}

We  prove the following stronger version of
Theorem~\ref{thm_gevrey}; it is not known whether this holds for
the Navier-Stokes equations.

\begin{theorem} \label{thm_gevrey_2}
Let us assume that $u(0) = u^{in} \in H$ and the force $f\in
G^{p_1}_{\sigma_1}$ for some $\sigma_1, p_1 > 0$. Then, there exists
$T$, depending only on the initial data, such that for every $0 < p
\le \min \{p_1, \log_\lambda \frac{1 + \sqrt{5}}{2} \}$ the
following holds
\begin{enumerate}
    \item For every $t \in [0, T]$ the solution $u(t)$ of the
    equation (\ref{eq_abstract_model}) remains bounded in
    $G^{p}_{\psi(t)}$, for $\psi(t) = \min \{t, \sigma_1 \}$.
    \item For every $t > T$ there exists $\sigma > 0$ such that
    the solution $u(t)$ of the equation (\ref{eq_abstract_model})
    remains bounded in $G^{p}_\sigma$.
\end{enumerate}
\end{theorem}

Before proving the theorem, we obtain the following estimates
for the nonlinear term:

\begin{lemma} \label{lem_gevrey_new}
Let $u, v\in G^{p}_\tau$, $w\in G^{p, 1}_\tau$ for some $\tau > 0$
and $p \le \log_\lambda \frac{1 + \sqrt{5}}{2}$. Then $B(u, v)$
belongs to $G^p_\tau$ and satisfies
\begin{equation} \label{eq_nl_gevrey_new}
\abs{(e^{\tau A^{p/2}} B(u, v), e^{\tau A^{p/2}} w)} \le C_6
\nm{u}_{G^{p}_\tau} \nm{v}_{G^{p}_\tau} \nm{w}_{G^{p, 1}_\tau},
\end{equation}
for an appropriate constant $C_6 > 0$.
\end{lemma}

\begin{proof}
Let us fix an arbitrary $p \le \log_\lambda \frac{1 + \sqrt{5}}{2}$.
Then we note that the operator $e^{\tau A^{p/2}}$ is diagonal with
eigenvalues $e^{\tau \widetilde{k}_n}$, $n = 1, 2, \dots$, where
$\widetilde{k}_n = \widetilde{\lambda}^n$ for $\widetilde{\lambda}
\le \frac{1 + \sqrt{5}}{2}$. Hence,
\begin{align}
\abs{(e^{\tau A^{p/2}} & B(u, v), e^{\tau A^{p/2}} w)} \le \notag
\\ & \le \sum_{n=1}^\infty e^{2 \tau \widetilde{k}_n} \babs{ a
k_{n+1} u_{n+2} v_{n+1}^* w_n^* + \notag \\ & + b k_n u_{n+1} w_n^*
v_{n-1}^* - c k_{n-1} w_n^* v_{n-1} u_{n-2}} \le \notag \\ & \le
\sup_{n \ge 1} \abs{e^{\tau \widetilde{k_n}} k_n w_n }
\sum_{n=1}^\infty e^{\tau \widetilde{k}_n} \bigg( \babs{ a \lambda
u_{n+2} v_{n+1}^* } + \notag \\ & + \babs{ b u_{n+1} v_{n-1}^* } +
\babs{ c \lambda^{-1} v_{n-1} u_{n-2} } \bigg) \le \notag \\
& \le C_6 \nm{w}_{G^{p, 1}_\tau} \nm{u}_{G^{p}_\tau}
\nm{v}_{G^{p}_\tau},
\end{align}
where we can choose $C_6 = (\lambda^{} a + b + \lambda^{-1} c)$. The
last inequality is the result of the Cauchy-Schwartz inequality and
the fact, that $e^{\tau (\widetilde{\lambda}^n -
\widetilde{\lambda}^{n+2} - \widetilde{\lambda}^{n+1})}$, $e^{\tau
(\widetilde{\lambda}^n - \widetilde{\lambda}^{n+1} -
\widetilde{\lambda}^{n-1})}$ and $e^{\tau (\widetilde{\lambda}^n -
\widetilde{\lambda}^{n-1} - \widetilde{\lambda}^{n-2})}$ are less
than or equal to $1$ for all $\tau$ and the specific choice of
$\widetilde{\lambda}$.

\end{proof}

Now we are ready to prove the Theorem:

\begin{proof} (Of Theorem~\ref{thm_gevrey_2})
Let us take an arbitrary $0 < p \le \min \{p_1, \log_\lambda \frac{1
+ \sqrt{5}}{2} \}$ and denote $\varphi(t) = \min(t, \sigma_1)$. For
a fixed time $t$ take the scalar product of the equation
(\ref{eq_shell_model}) with $u(t)$ in $G^{p}_\varphi(t)$ to get
\begin{align} \label{eq_gevrey_scalar_product_new}
(u'(t), u(t))_{p, \varphi(t)} & + \nu \nm{u(t)}_{p, \varphi(t)}^2
= \notag \\ & = (f, u(t))_{p, \varphi(t)} - (B(u(t), u(t)),
u(t))_{p, \varphi(t)}.
\end{align}

The left hand side of the equation could be transformed in the
following way:
\begin{align} \label{eq_gevrey_left side_new}
(u'(t), & A u(t))_{p, \varphi(t)} = (e^{\varphi(t) A^{p/2}} u'(t),
e^{\varphi(t) A^{p/2}} u(t)) = \notag \\ & = ((e^{\varphi(t)
A^{p/2}} u(t))' - \varphi'(t) A^{p/2} e^{\varphi(t) A^{p/2}} u(t),
e^{\varphi(t) A^{p/2}} u(t)) = \notag \\ & = \half \ddt
\abs{u(t)}^2_{p, \varphi(t)} - \varphi'(t) (A^{p/2} u(t),
u(t))_{p, \varphi(t)} \ge \notag \\ & \ge \half \ddt
\abs{u(t)}^2_{p, \varphi(t)} - \nm{u(t)}_{p, \varphi(t)}
\abs{u(t)}_{p, \varphi(t)} \ge \notag \\ & \ge \half \ddt
\abs{u(t)}^2_{p, \varphi(t)} - \frac{\nu}{4} \nm{u(t)}_{p,
\varphi(t)}^2 - \frac{1}{\nu} \abs{u(t)}_{p, \varphi(t)}^2.
\end{align}

The right hand side of the equation
(\ref{eq_gevrey_scalar_product_new}) could be bounded from above
in the usual way using Lemma~\ref{lem_gevrey_new}, Cauchy-Schwartz
and Young inequalities
\begin{align} \label{eq_gevrey_right_side_new}
\abs{(f, u(t))_{p, \varphi(t)} & - (B(u(t), u(t)), u(t))_{p,
\varphi(t)}} \le \notag \\ & \le \abs{f}_{p, \varphi(t)}
\abs{u(t)}_{p, \varphi(t)} + C_6 \nm{u}_{p, \varphi(t)}
\abs{u}_{p, \varphi(t)}^2 \le \notag \\ & \le \frac{3}{4}
\abs{f}_{p, \varphi(t)}^{4/3} + \frac{1}{4} \abs{u(t)}_{p,
\varphi(t)}^4 + \frac{\nu}{4} \nm{u}_{p, \varphi(t)}^2 +
\frac{C_6^2}{\nu} \abs{u}_{p, \varphi(t)}^4.
\end{align}

Combining (\ref{eq_gevrey_left side_new}) and
(\ref{eq_gevrey_right_side_new}), the equation
(\ref{eq_gevrey_scalar_product_new}) takes the form
\begin{align*}
\ddt \abs{u(t)}^2_{p, \varphi(t)} & + \nu \nm{A u(t)}_{p,
\varphi(t)}^2 \le \\ & \le \frac{3}{2} \abs{f}_{p, \varphi(t)}^{4/3}
+ \frac{2}{\nu} \abs{u(t)}_{p, \varphi(t)}^2 + (\half + \frac{2
C_6^2}{\nu}) \abs{u}_{p, \varphi(t)}^4 \le \\ & \le \frac{3}{2}
\abs{f}_{p, \varphi(t)}^{4/3} + C_7 \abs{u}_{p, \varphi(t)}^4,
\end{align*}
for some positive constant $C_7$.

Denoting $y(t) = 1 + \abs{u(t)}_{p, \varphi(t)}^2$ and $K =
\frac{3}{2} \abs{f}_{p, \varphi(t)}^{4/3} + C_7$ we get an
inequality
\[
\dot{y} \le K y^{2},
\]
It implies
\[
y(t) = 1 + \nm{u(t)}_{p, \varphi(t)}^2 \le 2 y(0) = 2(1 +
\abs{u^{in}}^2),
\]
for
\[
t \le T_1(\sigma_1, \abs{u^{in}}, \abs{f}_{p, \varphi(t)}) =
\frac{1}{2K} y^{-1}(0) = \frac{1}{2K} (1 + \abs{u^{in}}^2)^{-1}.
\]

According to Theorem~\ref{thm_weak_solution} the solution of the
equation (\ref{eq_abstract_model}) remains bounded in $H$ for all
time, if we start from $u^{in}\in H$. Moreover, from the
estimate (\ref{eq_u_bounds_weak_cont}), it follows that there
exists a constant $M$, such that for all $t > 0$
\[
\abs{u(t)} \le M.
\]
Hence we can repeat the above arguments starting from any time $t
> 0$ and find that
\begin{equation} \label{eq_gevrey_last_new}
\abs{e^{\varphi(T_2) A^{p/2}} A^{1/2} u(t)} \le 2 + 2 M^2,
\end{equation}
for all $t\ge T_2 = \frac{1}{2K} (1 + M^2)^{-1}$, and the Theorem
holds.

\end{proof}

\section{Global Attractors and Their Dimensions}

The first mathematical concept which we use to establish the
finite dimensional long-term behavior of the viscous sabra shell model is
the global attractor. The global attractor, $\mathcal{A}$, is the maximal
bounded invariant subset of the space $H$. It
encompasses all of the possible permanent regimes of the dynamics of
the  shell
model. It is also a compact subset of the space $H$
 which attracts all the trajectories of the system.
 Establishment of finite
Hausdorff and fractal dimensionality of the global attractor
implies the possible parameterization of the permanent regimes of
the dynamics in terms of a finite number of parameters. For the
definition and further discussion of the concept of the global
attractor see, e.g.,  \cite{FMRT01} and \cite{Te88}.

In analogy with Kolmogorov's mean rate of dissipation of energy in
turbulent flow we define
\[
\epsilon = \nu \ang{\nm{u}^2},
\]
the mean rate of dissipation of energy in the shell model system.
$\ang{\cdot}$ represents the ensemble average or a long-time
average. Since we do not know whether such long-time averages
converge for trajectories we will replace the above definition of
$\epsilon$ by
\[
\epsilon = \nu \sup_{u^{in}\in \mathcal{A}} \limsup_{t\to \infty}
\frac{1}{t} \int_0^t \nm{u(s)}^2 ds.
\]
We will also define the viscous dissipation length scale $l_d$.
According to Kolmogorov's theory, it should only depend on the
viscosity $\nu$ and the mean rate of energy dissipation
$\epsilon$. Hence, pure dimensional arguments lead to the
definition
\[
l_d = \bigg( \frac{\nu^{3}}{\epsilon} \bigg)^{1/4},
\]
which represents the largest spatial scale at which the viscosity
term begins to dominate over the nonlinear inertial term of the
shell model equation. In analogy with the conventional theory of
turbulence this is also the smallest scale that one needs to
resolve in order to get the full resolution for turbulent flow
associated with the shell model system.

We would like to obtain an estimate of the fractal
dimension of the global attractor for the system
(\ref{eq_abstract_model}) in terms of another non-dimensional
quantity -- the generalized Grashoff number. Suppose the forcing
term satisfies $f\in \l{}{\infty}([0, T], H)$ and denote
$\nm{f}_{\l{}{\infty}([0, \infty), H)} = \abs{f}_\infty$. Then we
define the generalized Grashoff number for our system to be
\begin{equation} \label{eq_grashof}
G = \frac{\abs{f}_\infty}{\nu^2 k_1^3}.
\end{equation}
The generalized Grashoff number was first introduced in the context
of the study of the finite dimensionality of long-term behavior of
turbulent flow in \cite{FMTT83}. To check that it is indeed
non-dimensional we note that $\nm{f}_{\l{}{\infty}([0, T], H)}$ has
the dimension of $\frac{L}{T^2}$, where $L$ is a length scale and
$T$ is a time scale. $k_1$ has the dimension of $\frac{1}{L}$ and
the kinematic viscosity $\nu$ has the dimension of $\frac{L^2}{T}$.
In order to obtain an estimate of the generalized Grashoff number,
which will be used in the proof of the main result of this section,
we can apply inequality (\ref{eq_strong_2}) to get
\begin{equation}\label{eq_grashof_bound}
\limsup_{t\to \infty} \frac{1}{t} \int_0^t \nm{u(s)}^2 ds \le
\frac{\abs{f}_\infty^2}{\nu^2 k_1^2} = \nu^2 k_1^4 G^2.
\end{equation}

\begin{theorem}
The Hausdorff and fractal dimensions of the global attractor of
the system of equations (\ref{eq_abstract_model}), $d_H(A)$ and
$d_F(A)$ respectively, satisfy
\begin{equation} \label{eq_dim_attractor}
d_H(A) \le d_F(A) \le \log_\lambda \bigg( \frac{L}{l_d} \bigg) +
\half \log_\lambda \bigg( C_1 (\lambda^2 - 1) \bigg).
\end{equation}

In terms of the Grashoff number $G$ the upper bound takes the form
\begin{equation} \label{eq_dim_attractor_grashof}
d_H(A) \le d_F(A) \le \log_\lambda G^{1/2} + \half \log_\lambda
\bigg( C_1 (\lambda^2 - 1) \bigg).
\end{equation}
\end{theorem}

\begin{proof}
We follow \cite{CF85} (see also \cite{ChepIl04_Attractor},
\cite{CF88}, \cite{Te88}) and linearize the shell model system of
equations (\ref{eq_abstract_model}) about the trajectory $u(t)$ in
the global attractor. In the Appendix~\ref{app_diff} we show that
the solution $u(t)$ is differentiable with respect to the initial
data, hence the resulting linear equation takes the form
\begin{subequations}
\label{eq__attr_lin}
\begin{gather}
\pdt{U} + \nu A U + B_0(t) U = 0 \label{eq_attr_lin_eq} \\
U(0) = U^{in}, \label{eq_attr_lin_initial_cond}
\end{gather}
\end{subequations}
where $B_0(t) U = B(u(t), U(t)) + B(U(t), u(t))$ is a linear
operator. In order to simplify the notation, we will denote
\[
\Lambda(t) = - \nu A - B_0(t).
\]

Let $U_j(t)$ be solutions of the above system satisfying $U_j(0) =
U_j^{in}$, for $j = 1, 2, \dots, N$. Assume now that $U_1^{in},
U_2^{in}, \dots, U_N^{in}$ are linearly independent in $H$, and
consider $Q_N(t)$ -- an $H$-orthogonal projection onto the span
$\{ U_j(t) \}_{j=1}^N$. Let $\{ \varphi_j(t) \}_{j=1}^N$ be the
orthonormal basis of the span of $\{ U_j(t) \}_{j=1}^N$. Notice
that $\varphi_j\in D(A)$ since $span \{ U_1(t), \dots, U_N(t) \}
\subset D(A)$.

Using the definition of $\Lambda(t)$ and the inequalities of the
Proposition~\ref{prop_non_linear} we get
\begin{align}
Re (Trace[\Lambda(t) \circ Q_N(t)]) & = Re \sum_{j=1}^N
(\Lambda(t) \varphi_j, \varphi_j) = \notag \\ & = \sum_{j=1}^N -
\nu \nm{\varphi_j}^2 - Re (B(\varphi_j, u(t)), \varphi_j) \le \notag \\
& \le - \nu \sum_{j=1}^N (k_j^2 + \abs{(B(\varphi_j, u(t)),
\varphi_j)}) \le \notag
\\ & \le - \nu k_0^2 \lambda^2 \frac{\lambda^{2N} - 1}{\lambda^2 -
1} + C_1 N \nm{u(t)}. \label{eq_trace_bound}
\end{align}

According to the recent results of \cite{ChepIl04_Attractor} (see
also \cite{CF88}, \cite{CF85}, \cite{Te88}), if $N$ is large enough,
such that
\[
\limsup_{t\to \infty} \frac{1}{t} \int_0^t Re (Trace[\Lambda(t)
\circ Q_N(s)]) ds < 0,
\]
then the fractal dimension of the global attractor is bounded by
$N$. We need therefore to estimate $N$ in terms of the energy
dissipation rate. Using (\ref{eq_trace_bound}), the definition of
$\epsilon$ we find that it is sufficient to require $N$ to be large
enough such that it satisfies
\begin{equation} \label{eq_attr_ineq_nu}
\nu k_0^2 \lambda^2 \frac{\lambda^{2N} - 1}{\lambda^2 - 1} > C_1 N
\bigg( \limsup_{t\to \infty} \frac{1}{t} \int_0^t \nm{u(t)}^2
\bigg)^{1/2} = C_1 N \bigg( \frac{\epsilon}{\nu} \bigg)^{1/2}.
\end{equation}
Finally, $\epsilon \nu^{-3} = l_d^{-4}$ implies
\[
\lambda^{2N} > \frac{\lambda^{2N} - 1}{N} > C_1 \frac{\lambda^2 -
1}{k_1^2 l_d^2} = C_1 (\lambda^2 - 1) \bigg( \frac{L}{l_d} \bigg)^2,
\]
which proves (\ref{eq_dim_attractor}). Applying the estimate
(\ref{eq_grashof_bound}) to the inequality (\ref{eq_trace_bound}) we
get the bound (\ref{eq_dim_attractor_grashof}) in terms of the
generalized Grashoff number.

\end{proof}

\section{Determining Modes}


Let us consider two solutions of the the shell model equations $u,
v$ corresponding to the forces $f, g\in \l{}{2}([0, \infty), H)$
\begin{eqnarray}
  \frac{du}{dt} + \nu Au + B(u, u) = f, \label{eq_dm_eq1} \\
  \frac{dv}{dt} + \nu Av + B(v, v) = g. \label{eq_dm_eq2}
\end{eqnarray}

We give a slightly more general definition of a notion of the
determining modes, than the one that is introduced previously in
literature \cite{FP67} (see also \cite{FMRT01}, \cite{FMTT83},
\cite{JT92} and references therein).
\begin{definition}
We define a set of determining modes as a finite set of indices
$\mathcal{M}\subset \mathbb{N}$, such that whenever the forces $f,
g$ satisfy
\begin{equation} \label{eq_dm_force}
\abs{f(t) - g(t)} \to 0, \;\;\; \text{as} \;\; t\to \infty,
\end{equation}
and
\begin{equation}
\sum_{n\in \mathcal{M}} \abs{u_n(t) - v_n(t)}^2 \to 0, \;\;\;
\text{as} \;\; t\to \infty
\end{equation}
it follows that
\begin{equation}
\abs{u(t) - v(t)} \to 0, \;\;\; \text{as} \;\; t\to \infty.
\end{equation}
The number of determining modes $N$ of the equation is the size of
the smallest such a set $\mathcal{M}$.
\end{definition}

We would like to recall the following generalization of the
classical Gronwall's lemma which was proved in \cite{JT92},
\cite{JT93_Detmodesnodes} (see also, \cite{FMRT01}).

\begin{lemma} \label{lem_dm}
Let $\alpha = \alpha(t)$ and $\beta = \beta(t)$ be locally
integrable real-valued functions on $[0, \infty)$ that satisfy the
following condition for some $T > 0$:
\begin{eqnarray}
  \liminf_{t\to \infty} \frac{1}{T} \int_t^{t+T} \alpha(\tau) d\tau & > & 0,
\label{eq_dm_lem_1} \\
  \limsup_{t\to \infty} \frac{1}{T} \int_t^{t+T} \alpha^-(\tau) d\tau & < &
\infty, \label{eq_dm_lem_2} \\
  \lim_{t\to \infty} \frac{1}{T} \int_t^{t+T} \beta^+(\tau) d\tau & = &
  0, \label{eq_dm_lem_3}
\end{eqnarray}
where $\alpha^-(t) = \max\{-\alpha(t), 0\}$ and $\beta^+(t) =
\max\{\beta(t), 0\}$. Suppose that $\xi = \xi(t)$ is an absolutely
continuous nonnegative  function on $[0, \infty)$ that satisfies the
following inequality almost everywhere on $[0, \infty)$:
\begin{equation}
\frac{d\xi}{dt} + \alpha \xi \le \beta.
\end{equation}
Then $\xi(t)\to 0$ as $t\to \infty$.
\end{lemma}

A weaker version of the above statement was introduced in
\cite{FMTT83}. In fact this weaker version would be sufficient for
our purposes.

\begin{theorem} \label{thm_det_modes}
Let $G$ be a Grashoff number defined in (\ref{eq_grashof}). Then,
\begin{enumerate}
    \item The first $N$ modes are determining modes for the shell model equation
    provided
    \begin{equation} \label{eq_dm_size}
    N > \half \log_\lambda (C_1 G).
    \end{equation}

    \item Let $u, v$ be two solutions of the equations
    (\ref{eq_dm_eq1}) and (\ref{eq_dm_eq2}) correspondingly. Let
    the forces $f, g$ satisfy (\ref{eq_dm_force}) and integer $N$ be defined
    as in (\ref{eq_dm_size}). If
    \[
    \lim_{t\to \infty} \abs{u_N(t) - v_N(t)}, \;\;\;
    \text{and} \;\;\; \lim_{t\to \infty} \abs{u_{N-1}(t) - v_{N-1}(t)} =
    0,
    \]
    then
    \[
    \lim_{t\to \infty} \abs{Q_N u(t) - Q_N v(t)} = 0.
    \]
\end{enumerate}
\end{theorem}

\begin{proof}
Denote $w = u - v$, which satisfies the equation
\begin{equation}\label{eq_dm_diff_equation}
\frac{dw}{dt} + \nu Aw + B(u, w) + B(w, v) = f - g.
\end{equation}
Let us fix an integer $m > 1$ and define $P_m : H \to H$ to be an
orthogonal projection onto the first $m$ coordinates, namely, onto
the subspace of $H$ spanned by $m$ vectors of the standard basis $\{
\phi_i \}_{i=1}^m$. Also denote $Q_m = I - P_m$.

It is not hard to see that the second part of the Theorem implies
the first statement. Therefore it is enough for us to prove that if
\[
\lim_{t\to \infty} \abs{w_N(t)}, \;\;\; \text{and} \;\;\;
\lim_{t\to \infty} \abs{w_{N-1}(t)} = 0,
\]
then for every $m > \half \log_\lambda (C_1 G)$
\[
\lim_{t\to \infty} \abs{Q_m w} = 0.
\]

Multiplying the equation (\ref{eq_dm_diff_equation}) by $Q_m w$ in
the scalar product of $H$ we get
\begin{align}
\half \ddt \abs{Q_m w}^2 & + \nu \nm{Q_m w}^2 = \notag \\ & = Re (f
- g, Q_m w) - Re (B(u, w), Q_m w) - Re (B(w, v), Q_m w).
\end{align}
For the left hand side of the equation we can use the inequality
$k_{m+1}^2 \abs{Q_m w} \le \nm{Q_m w}$. For the right hand side of
the equation we observe that
\begin{align*}
Re & (B(u, w), Q_m w) = Im \sum_{n = m + 1}^\infty \bigg( a
k_{n+1} w_{n+2} u_{n+1}^* w_{n}^* + b k_n w_{n+1} w_{n}^*
u_{n-1}^* + \\ & + a k_{n-1} w_{n}^* u_{n-1} w_{n-2} + b k_{n-1}
w_{n}^* w_{n-1} u_{n-2} \bigg) \le \\ & \le \babs{ a k_m w_{m+1}^*
u_m w_{m-1} + a k_{m+1} w_{m+2}^* u_{m+1} w_{m} + b k_{m}
w_{m+1}^* w_m u_{m-1} }.
\end{align*}
Moreover,
\begin{align*}
Re & (B(w, v), Q_m w) = Im \sum_{n = m + 1}^\infty \bigg( a
k_{n+1} v_{n+2} w_{n+1}^* w_{n}^* + b k_n v_{n+1} w_{n}^*
w_{n-1}^* + \\ & + a k_{n-1} w_{n}^* w_{n-1} v_{n-2} + b k_{n-1}
w_{n}^* v_{n-1} w_{n-2} \bigg) \le \\ & \le \babs{ a k_m w_{m+1}^*
w_m v_{m-1} + a k_{m+1} w_{m+2}^* w_{m+1} v_{m} + \\ & + b k_{m}
w_{m+1}^* v_m w_{m-1} } + C_{1} \nm{v} \abs{Q_m w}^2.
\end{align*}

Finally denoting $\xi = \abs{Q_m w}^2$ we can rewrite the equation
in the form
\[
\half \frac{d \xi}{dt} + \alpha \xi \le \beta,
\]
where
\[
\alpha(t) = \nu k_{m+1}^2 - C_{1} \nm{v(t)},
\]
and
\begin{align*}
\beta(t) & = (f - g, Q_m w) + \\ & + \babs{ a k_m w_{m+1}^* u_m
w_{m-1} + a k_{m+1} w_{m+2}^* u_{m+1} w_{m} + b k_{m} w_{m+1}^*
w_m u_{m-1} } + \\ & + \babs{ a k_m w_{m+1}^* w_m v_{m-1} + a
k_{m+1} w_{m+2}^* w_{m+1} v_{m} + b k_{m} w_{m+1}^* v_m w_{m-1} }.
\end{align*}
To see that $\beta(t)$ satisfies condition (\ref{eq_dm_lem_3}) of
Lemma~\ref{lem_dm} it is enough to show that $\beta(t) \to 0$ as
$t\to \infty$. It is clearly true, if we assume that the forces
$f, g$ satisfy (\ref{eq_dm_force}) and $\abs{w_m(t)},
\abs{w_{m-1}(t)} \to 0$ as $t\to \infty$. Moreover, it is easy to
see that $\alpha(t)$ satisfies the condition (\ref{eq_dm_lem_2}).
Hence, in order to apply Lemma~\ref{lem_dm} we need to show that
$\alpha(t)$ also satisfies the conditions (\ref{eq_dm_lem_1}),
namely, that
\begin{equation} \label{eq_detmodes_ineq_nu}
\liminf_{t\to \infty} \frac{1}{T} \int_t^{t+T} \alpha(\tau) d\tau
\ge \nu k_{m+1}^2 - C_1 \bigg( \limsup_{t\to \infty} \frac{1}{T}
\int_t^{t+T} \nm{v(\tau)}^2 d\tau \bigg)^{1/2} > 0.
\end{equation}
To estimate the last quantity, we can use the bound
(\ref{eq_grashof_bound}) to conclude that if $m$ satisfies
\[
\lambda^{2m} > C_1 G,
\]
then
\[
\abs{Q_m w} \to 0 \;\;\; \text{as} \;\; t\to \infty.
\]

\end{proof}

\begin{remark} \label{rem_det_modes}
The existence of the determining modes for the Navier-Stokes
equations both in two and three dimensions is known (see,
e.g., \cite{CFMT85}
\cite{FMRT01}, \cite{JT93_Detmodesnodes} and
reference therein). However, there exists a gap
between the upper bounds for the lowest number of determining
modes and the dimension of the global attractor for the
two-dimensional Navier-Stokes equations both for the no-slip and
periodic boundary conditions. Our upper bounds for the dimension
of the global attractor and for the number of determining modes
for the sabra shell model equation coincides. Recently it was
shown in \cite{IT05_Attractor} that a similar result is
also true for
the damped-driven NSE and the Stommel-Charney
barotropic model of
ocean circulation.
\end{remark}

\section{Existence of Inertial Manifolds}

In this section we prove the existence of a finite-dimensional
inertial manifold (IM). The concept of inertial manifold for
nonlinear evolution equations was first introduced in
\cite{FSTe88} (see also, e.g., \cite{CF88}, \cite{ConstFoNiTe88_Inertial},
\cite{ConstFoNiTe89_Inertial2}, \cite{FST89},
\cite{SellYou92_Inertial}, \cite{Te88}). An inertial
manifold is a finite
dimensional Lipschitz, globally invariant manifold which attracts
all bounded sets in the phase space at an exponential rate and,
consequently,  contains the global attractor. In fact, one can show that
the IM is smoother, in particular $C^1$ (see, e.g.,
\cite{DemGhi91_Inertial}, \cite{Sri-Ou},
\cite{SellYou92_Inertial}).
The smoothness  and
invariance under the reduced dynamics of the IM implies that a finite
system of ordinary differential equations is equivalent to
the original infinite system. This is the ultimate and  best
notion of system reduction that one could hope for. In other
words, an IM is an exact rule for parameterizing the large modes
(infinite many of them) in terms of the low ones (finitely many of
them).

In this section we use Theorem $3.1$ of \cite{FST89} to show the
existence of inertial manifolds for the system
(\ref{eq_abstract_model}) and to estimate its dimension. Let us
state Theorem $3.1$ of \cite{FST89} in the following way (see also
\cite{Te88}, Chapter VIII, Theorem $3.1$)

\begin{theorem} \label{thm_inertial}
Let the nonlinear term of the equation (\ref{eq_inertial_manifold})
$R(\cdot)$ be a differentiable map from $D(A)$ into $D(A^{1 -
\beta})$ satisfying
\begin{subequations}
\label{eq_im_non_linear_term}
\begin{gather}
\abs{R'(u) v} \le C_8 \abs{A u} \abs{A^{\beta} v}
\label{eq_im_non_linear_term1}
\\
\abs{A^{1 - \beta} R'(u) v} \le C_9 \abs{A u} \abs{A v},
\label{eq_im_non_linear_term2}
\end{gather}
\end{subequations}
for some $0 \le \beta < 1$, for all $u, v\in D(A)$ and appropriate
constants $C_8, C_9 > 0$, which depend on physical parameters $a, b,
c, \nu$ and $f$.

Let $\tilde{b} >0$  and $l \in [0,1)$ be two fixed numbers. Assume
that there exists an $N$, large enough, such that the eigenvalues of
$A$ satisfy
\begin{equation} \label{eq_im_spectral_gap}
\frac{k_{N+1} - k_N}{{k_{N+1}^\beta + k_N^\beta}} \ge \max \bigg \{
\frac{2}{l} (1 + l) K_2, \bigg( \frac{1}{1 - \beta} \bigg)
\frac{\tilde{b}}{K_1} \bigg \}
\end{equation}
where
\[
K_1 = \abs{A^{1-\beta} f} + 4 C_9 \rho^2,
\]
\[
K_2 = \frac{8}{\rho} \abs{A^{1-\beta} f} + 26 C_9 \rho,
\]
$\rho$ is the radius of the absorbing ball in $D(A)$, whose
existence is provided by Proposition~\ref{prop_absorbing_balls}.

Then the equation (\ref{eq_inertial_manifold}) possesses an inertial
manifold of dimension $N$ which  is a graph of a function $\Phi: P_N
H \to Q_N D(A)$ with

\[
\tilde{b} = \sup_{\{p\in P_N H\}} |A\Phi(p)|,
\]
and
\[
l =\sup_{\{p_1,p_2\in P_N H, p_1\neq p_2\}}
\frac{|A(\Phi(p_1)-\Phi(p_2))|}{|A(p_1-p_2)|}, \]
 is the Lipschitz constant of $\Phi$.
\end{theorem}

In order to apply the Theorem, we need to show that the sabra
model equation satisfies the properties of the abstract framework
of \cite{FST89}. Let us rewrite our system
(\ref{eq_abstract_model}) in the form
\begin{equation} \label{eq_inertial_manifold}
\pdt{u} + \nu Au + R(u) = 0,
\end{equation}
where the nonlinear term $R(u) = B(u, u) - f$.

First of all, we need to show that the nonlinear term $R(u)$ of the
equation (\ref{eq_inertial_manifold}) satisfies
(\ref{eq_im_non_linear_term}) for $\beta = 0$. Indeed, and based on
(\ref{eq_nl_3}) $R(u)$ is a differentiable map from $D(A)$ to
$D(A^{1/2}) = V$, where
\[
R'(u) w = B(u, w) + B(w, u), \;\;\; \forall u, w\in D(A).
\]
Moreover, the estimates (\ref{eq_im_non_linear_term}) are
satisfied according to Proposition \ref{prop_non_linear}, namely
\[
\abs{R'(u) v} = \abs{B(u, v) + B(v, u)} \le (C_1+C_2) \nm{u}\abs{v}
\le \frac{(C_1+C_2)}{k_1}\abs{Au}\abs{v},
\]
 where the last inequality results from the fact that $k_1 \nm{u}
\le \abs{A u}$ and the constant $C_8 = \frac{(C_1+C_2)}{k_1}
=\frac{(\abs{a} (\lambda^{-1} + \lambda) + \abs{b} (\lambda^{-1} +
1)) + (2 \abs{a} + 2 \lambda \abs{b})}{k_1}$. In addition, one can
prove as in Proposition \ref{prop_non_linear} that
\[
\abs{AR'(u) v} \le \abs{AB(u, v)} + \abs{AB(v, u)} \le C_9' \nm{u}
\abs{A v} \le C_9 \abs{A u} \abs{A v},
\]
where here again the last inequality results from the fact that $k_1
\nm{u} \le \abs{A u}$ and the constant  $C_9 = C_9' / k_1$.

Finally, because of the form of the wavenumbers $k_n$
(\ref{eq_freq}), the spectral gap condition
(\ref{eq_im_spectral_gap}) is satisfied for $\beta=0$. Therefore, we
can apply the Theorem~\ref{thm_inertial} to equation
(\ref{eq_abstract_model}), and conclude that the sabra shell model
possesses an inertial manifold.

\subsection{Dimension of Inertial Manifolds}

Let us calculate the dimension of the inertial manifold of the the
sabra shell model equation for the specific choice of parameters
\[
\beta = 0, \;\;\; l = \half, \;\;\; \tilde{b} = \rho,
\]
where $\rho$ is the radius of the absorbing ball in the norm of the
space $D(A)$ (see Proposition~\ref{prop_absorbing_balls}). In that
case, the conditions (\ref{eq_im_spectral_gap}) take the form
\begin{equation} \label{eq_im_spectral_gap_new}
k_{N+1} - k_N \ge \max \bigg \{ 4 K_2, \frac{2 \rho}{K_1} \bigg \},
\end{equation}
and
\begin{corollary}
The equation (\ref{eq_abstract_model}) possesses an inertial
manifold of dimension
\begin{equation} \label{eq_inertial_manifold_N}
N \ge \max \bigg \{ \log_\lambda \frac{4 K_2}{\lambda - 1},
\log_\lambda \frac{2 \rho}{(\lambda - 1) K_1} \bigg \}.
\end{equation}
\end{corollary}

\noindent
\section{Conclusions}

We have established the global regularity of the sabra shell model
of  turbulence. We shown analytically  that the shell model enjoys
some of the commonly observed features of real world turbulent
flows. Specifically,  we have established using the Gevrey
regularity technique  the existence of an exponentially decaying
dissipation range, which is consistent with the observations
established in \cite{SchKaLo95} for the GOY model using asymptotic
methods. Moreover, we have provided explicit upper bounds, in terms
of the given parameters of the shell model, for the number
asymptotic degrees of freedom. Namely, we presented explicit
estimates for the dimension of the global attractor and for the
number of determining modes. In the forthcoming paper we will
investigate the lower bounds on the Hausdorff dimension of the
global attractor.

Finally, we have shown that the shell model possess  a finite
dimensional inertial manifold. That is, there exists an exact rule
which parameterizes the small scales as a function of the large
scales. The existence of inertial manifolds is not known for the
Navier-Stokes equations. It is worth mentioning that the tools
presented here can be equally applied to other shell models.

\noindent
\section*{Acknowledgments}

The authors would like to thank Professors I. Procaccia, V. Lvov and
Dr. A. Pomyalov for the very stimulating and inspiring discussions.
The work of P.C. was partially supported by the NSF grant No.
DMS-0504213. The work of E.S.T. was supported in part by the NSF
grants No. DMS--0204794 and DMS--0504619,  the MAOF Fellowship of
the Israeli Council of Higher Education, and by the USA Department
of Energy, under contract number W--7405--ENG--36 and ASCR Program
in Applied Mathematical Sciences.

\appendix

\newpage
\section{Dissipativity in $D(A)$} \label{app_abs_balls}

In this section we prove that the equation (\ref{eq_abstract_model})
is dissipative in different norms and that its solutions are real
analytic with respect to the time variable with values in $D(A)$.

First, we start with the space $H$. Consider the relation
(\ref{eq_strong_4}), and by letting $t\to \infty$ we obtain
\begin{equation} \label{eq_radius_diss_ball_h}
\limsup_{t\to \infty} \abs{u(t)} \le \frac{\abs{f}_\infty}{\nu
k_1^2} = G \nu k_1,
\end{equation}
where $\abs{f}_\infty = \nm{f}_{\l{}{\infty}([0, \infty), H)}$ and
$G$ is the generalized generalized Grashoff number, defined in
(\ref{eq_grashof}).

The existence of the absorbing ball in the space $V$ readily
follows from Theorem~\ref{thm_strong}. However, the exact formula
for the radius of the absorbing ball is much more involved.

We already mentioned that if we assume that the forcing $f$ belongs
to some Gevrey class for all $t\ge 0$, then the existence of
absorbing balls for solutions of the equation
(\ref{eq_abstract_model}) in the norms of $V$ and $D(A)$ follows
from the fact that in that case all the functions belonging to the
global attractor are in some Gevrey class and hence are bounded in
all weaker norms. In the current section we will show that the
strong assumption on the force $f$ to be in the Gevrey class could
be dropped. In particular, we will assume that $f\in H$ is time
independent. Our proof follows \cite{FTe79} (see also \cite{CF88},
ch. 12).

Fix $m\in \N$, let $P_m$ be the projection of $H$ onto the first
$m$ coordinates and denote
\[
P_m u = u^m.
\]
First, let us complexify our equation and all the relevant spaces
and operators. Recall that the Galerkin approximating system
(\ref{eq_abstract_model_galerkin}) of order $m$ for equation
(\ref{eq_abstract_model}) is an $m$ - dimensional system of ordinary
differential equations, with analytic nonlinearity, and hence its
solutions are locally complex analytic functions.

Let us consider the complex time variable $t = s e^{i \theta}$,
$\theta\in (- \frac{\pi}{2}, \frac{\pi}{2})$ and $s\in \R_+$. Then
\begin{align*}
\frac{d}{ds} \nm{u^m(s e^{i \theta})}^2 = \frac{d}{ds} (u^m(s e^{i
\theta}), A u^m(s e^{i \theta})) = 2 Re (e^{i\theta} \ddt u^m, A
u^m).
\end{align*}
Then, by taking the scalar product of (\ref{eq_galerkin}) with $A
u^m$, multiplying by $e^{i\theta}$ and taking the real part we obtain
\begin{align*}
\half \frac{d}{ds} \nm{u^m(s e^{i \theta})}^2 & + \nu \cos \theta
\abs{A u^m(s e^{i \theta})}^2 = \\ & = Re \bigg( e^{i \theta} (f,
A u^m) - e^{i \theta} (P_m B(u^m, u^m), A u^m) \bigg) \le \\ & \le
\frac{\nu \cos \theta}{2} \abs{A u^m(s e^{i \theta})}^2 +
\frac{\abs{f}^2}{\nu \cos \theta} + \frac{C_1^2}{\nu \cos \theta}
\nm{u^m(s e^{i \theta})}^4,
\end{align*}
where the last line follows from (\ref{eq_nl_vdnorm}) and Young's
inequality. Finally we deduce
\[
\frac{d}{ds} \nm{u^m(s e^{i \theta})}^2 + \nu \cos \theta \abs{A
u^m(s e^{i \theta})}^2 \le \frac{2 \abs{f}^2}{\nu \cos \theta} +
\frac{2 C_1^2}{\nu \cos \theta} \nm{u^m(s e^{i \theta})}^4.
\]

Now we are able to derive the bound
\begin{equation}\label{eq_a_bound}
\nm{u^m(t)}^2 \le 2 (\nm{u^m(0)}^2 + 1) \le 2 (\nm{u^{in}}^2 + 1),
\end{equation}
provided $t = s e^{i\theta}$ satisfies
\begin{equation}\label{eq_a_time_bound}
s \le \frac{1}{2 K} (\nm{u^{in}}^2 + 1)^{-1},
\end{equation}
where $K = \frac{2 \abs{f}^2}{\nu \cos \theta} + \frac{2
C_1^2}{\nu \cos \theta}$.

The bounds (\ref{eq_a_bound}) and (\ref{eq_a_time_bound}) show that
for every $m\in \N$, the functions $u^m(t): \C \to \C^m$ are
analytic in the domain
\[
D = D(\nu, \nm{u^{in}}, \abs{f}) = \{ t = s e^{i\theta} \;:\;
\abs{\theta} < \frac{\pi}{2}, \; 0 < s \le \frac{1}{2 K}
(\nm{u^{in}}^2 + 1)^{-1} \}.
\]
Let $\gamma$ be a small circle contained in the domain $D$. Then
according to Cauchy formula
\[
\ddt u^m(t) = \frac{1}{2\pi i} \int_\gamma \frac{u^m(z)}{(z - t)^2}
dz.
\]
Using the relation (\ref{eq_a_bound}) we get
\[
\nm{\ddt u^m} \le \frac{\sqrt{2}}{r_\gamma} (\nm{u^m(0)}^2 +
1)^{1/2},
\]
where $r_\gamma$ is the radius of $\gamma$. Let $M\subset D$ be a
compact subset of $D$ and denote $r_M = dist(M, \partial D)$, then
\begin{equation} \label{eq_derivative_bound}
\nm{\ddt u^m} \le \frac{2\sqrt{2}}{r_M} (\nm{u^m(0)}^2 + 1)^{1/2},
\end{equation}
for all $t\in M$ and all $m\in \N$.

Finally, comparing the $H$ norms of both sides of the equation
(\ref{eq_galerkin}), using the relations (\ref{eq_vh}),
(\ref{eq_derivative_bound}) and the triangle inequality and we get
\begin{align*}
\nu \abs{A u^m(t)} & \le \abs{\ddt u^m} + \abs{f(t)} + \abs{B(u^m,
u^m)} \le
\\
& \le k_1^{-1/2} \nm{\ddt u^m} + \abs{f(t)} + \abs{B(u^m, u^m)} \le
\\
& \le \frac{2\sqrt{2}}{r_M\sqrt{k_1}} (\nm{u^m(0)}^2 + 1)^{1/2} +
\abs{f(t)} + C_1 \abs{u^m(t)} \nm{u^m(t)},
\end{align*}
for all $t\in M$. Since all the summands at the right hand side are
bounded (see the proofs of Theorem~\ref{thm_weak_solution} and
Theorem~\ref{thm_strong}) and those bounds do not depend on $m$ we
can pass to the limit in $m$, using Vitali's convergence theorm for
complex analytic functions, concluding that there exists a constant
$E = E\big(\nu, \abs{f}, \nm{u^{in}}, \lambda, k_0, a, b, c\big)$,
independent of $m$, such that
\[
\abs{A u(t)} \le E,
\]
for all $t\in M$. Moreover, for all $t\in M$ the solution $u(t)$ is
a complex analytic function with values in $D(A)$. Repeating this
procedure starting instead of $0$ from some $t_1\in M$ we conclude
that $\abs{A u(t)}$ is uniformly bounded for all $t > 0$.

\begin{proposition} \label{prop_absorbing_balls}
Let $f\in H$ and $u^{in}\in V$. Then the solution of the equation
(\ref{eq_abstract_model}) is an analytic function, with respect to
the time variable with values in $D(A)$, and that it possesses
absorbing balls in the spaces $H, V$ and $D(A)$.
\end{proposition}

\section{Differentiability of the semigroup
             with respect to initial conditions} \label{app_diff}

Theorems \ref{thm_weak_solution} and \ref{thm_cont_dependence}
shows in fact, that the initial-value problem
(\ref{eq_abstract_model}) is well posed. This allows us to define
the semigroup $S(t)$, i.e., the one parameter semigroup family of
operators
\[
S(t) : u^{in}\in H \to u(t)\in H,
\]
which are bounded for almost all $t \ge 0$. According to the
Theorem~\ref{thm_cont_dependence}, the mapping $S(t) : H \to H$ is
Lipschitz continuous with respect to the intial
data. The purpose of this section is to show that this mapping is
also Fr\'{e}chet differentiable with respect to the initial
data (see, e.g., \cite{Te88} (ch. VI, 8)).

First of all we wish to linearize the non-linear term. Let us fix $T
> 0$ and let $u, v\in \l{}{\infty}([0, T], H) \cap \l{}{2}([0, T],
V)$ be solutions of the equation (\ref{eq_abstract_model}) with the
initial conditions $u(0) = u^{in}$ and $v(0) = v^{in}$. Then for
almost every $t\in [0, T]$ we can write
\[
B(u(t), u(t)) - B(v(t), v(t)) = B_0(t) (u(t) - v(t)) + B_1(t, u -
v),
\]
where $B_0(t) : H \to V'$ is a linear operator defined as
\[
B_0(t) w = B(u(t), w(t)) + B(w(t), u(t)),
\]
and
\[
B_1(t, w) = - B(w(t), w(t)).
\]

Let us consider a solution $U(t)$ of the linearized equation
\begin{subequations}
\label{eq_lin}
\begin{gather}
\pdt{U} + \nu A U + B_0(t) U = 0 \label{eq_lin_eq} \\
U(0) = U^{in} = u^{in} - v^{in}, \label{eq_lin_initial_cond}
\end{gather}
\end{subequations}
satisfying $U\in \l{}{\infty}([0, T], H) \cap \l{}{2}([0, T], V)$.
Denote
\[
\varphi = u - v - U.
\]

Using the fact that $u, v$ satisfy equation
(\ref{eq_abstract_model}) and $U$ is the solution of
(\ref{eq_lin}) we can directly check that $\varphi$ satisfies
\begin{equation}
\pdt{\varphi} + \nu A \varphi + B_0(t) \varphi = - B_1(t, u(t) -
v(t)),
\end{equation}
and $\varphi(0) = 0$. Multiplying by $\varphi$ we get
\begin{equation}
\half \ddt \abs{\varphi}^2 + \nu \nm{\varphi}^2 + \ang{B_0(t)
\varphi, \varphi} = - \ang{B_1(t, u(t) - v(t)), \varphi}.
\end{equation}

According to the Proposition~\ref{prop_non_linear} we can rewrite
the last equation as the inequality
\[
\half \ddt \abs{\varphi}^2 + \nu \nm{\varphi}^2 \le C'_8
\abs{\varphi} \cdot \abs{u} \cdot \nm{\varphi} + C_9 \abs{u - v}^2
\cdot \nm{\varphi}.
\]

The fact that $u\in \l{}{\infty}([0, T], H)$ together with Young's
inequality imply
\[
\ddt \abs{\varphi}^2 \le \ddt \abs{\varphi}^2 + \nu \nm{\varphi}^2
\le \frac{C_8}{\nu} \abs{\varphi}^2 + \frac{C_9}{\nu} \abs{u -
v}^4.
\]
Our next step is to use Gronwall's inequality to get
\[
\abs{\varphi(t)}^2 \le \frac{C_9}{\nu} \int_0^T e^{\frac{C_8}{\nu}
(T - s)} \abs{u(s) - v(s)}^4 ds,
\]
for every $t\in (0, T]$. Once again applying
Theorem~\ref{thm_cont_dependence} we finally get
\[
\abs{\varphi(t)}^2 \le C_{10} \abs{u^{in} - v^{in}}^4,
\]
where $0 < C_{10} = \frac{C_9^2 e^{4K}}{\nu} \int_0^T
e^{\frac{C_8}{\nu} (T - s)} ds$ and $K$ is the constant from the
statement of the Theorem~\ref{thm_cont_dependence}. It follows
that
\[
\frac{\abs{u(t) - v(t) - U(t)}}{\abs{u^{in} - v^{in}}} \le
\abs{u^{in} - v^{in}} \longrightarrow 0, \;\;\; 0 < t \le T,
\]
as $\abs{u^{in} - v^{in}}$ tends to $0$. That exactly means that
$U(t)$ is a differential of $S(t)$ with respect to $u^{in}\in H$.


\begin{thebibliography}{99}


\bibitem{Au63}
  J. P. Aubin,
  \textit{Un theorem de compacite},
  C. R. Acad. Sci. Paris Ser. I Math. \textbf{256} (1963), 5042-5044.


\bibitem{Bi03}
  L. Biferale,
  \textit{Shell models of energy cascade in turbulence},
  Annual Rev. Fluid Mech., \textbf{35} (2003), 441-468.


\bibitem{BJPV98}
  T. Bohr, M. H. Jensen, G. Paladin, A. Vulpiani,
  \textit{Dynamical Systems Approach to Turbulence},
  Cambridge University press, 1998.


\bibitem{ChepIl04_Attractor}
  V. V. Chepyzhov, A. A. Ilyin,
  \textit{On the fractal dimension of invariant sets; Applications to Navier-Stokes equations},
  Disc. Cont. Syn. Systems, \textbf{10 (1\&2)} (2004), 117-135.


\bibitem{CF88}
  P. Constantin, C. Foias,
  \textit{Navier-Stokes Equations},
  The University of Chicago Press, 1988.


\bibitem{CF85}
  P. Constantin, C. Foias,
  \textit{Global Lyapunov exponents, Kaplan-Yorke formulas and the dimensional of the attractors for 2D Navier-Stokes equations},
  Comm. Pure Appl. Math, \textbf{38} (1985), 1-27.


\bibitem{CFMT85} P. Constantin, C. Foias, O. Manley, R. Temam,
  \textit{Determining modes and fractal dimension of turbulent flows},
  J. Fluid Mech., \textbf{150}, (1985), 427-440.


\bibitem{ConstFoNiTe88_Inertial}
  P. Constantin, C. Foias, B. Nicolaenko, R. Temam,
  \textit{Integral Manifolds and Inertial Manifolds for Dissipative Partial Differential Equations},
  Applied Mathematics Sciences, \textbf{70}, Springer-Verlag, 1988.


\bibitem{ConstFoNiTe89_Inertial2}
  P. Constantin, C. Foias, B. Nicolaenko, R. Temam,
  \textit{Spectral barriers and inertial manifolds for dissipative partial differential equations},
  J. Dynamics Differential Eqs., \textbf{1}, (1989), 45-73.


\bibitem{DemGhi91_Inertial}
  F. Demengel, J.-M. Ghidaglia,
  \textit{Some remarks on the smoothness of inertial manifolds},
  Nonlinear Analysis-TMA, \textbf{16} (1991), 79-87.


\bibitem{DT95_SpectrumDecay}
  C. R. Doering, E. S. Titi,
  \textit{Exponential decay-rate of the power spectrum for solutions of the
Navier-Stokes equations},
  Phys. of Fluids, \textbf{7 (6)} (1995), 1384-1390.


\bibitem{Gl73}
  E. B. Gledzer,
  \textit{System of hydrodynamic type admitting two quadratic
           integrals of motion},
  Sov. Phys. Dokl., \textbf{18} (1973), 216-217.


\bibitem{Fr95}
  U. Frisch,
  \textit{Turbulence: The Legacy of A. N. Kolmogorov},
  Cambridge University press, 1995.


\bibitem{FMRT01}
  C. Foias, O. Manley, R. Rosa, R. Temam,
  \textit{Navier-Stokes Equations and Turbulence},
  Cambridge University press, 2001.


\bibitem{FMTT83}
  C. Foias, O. Manley, R. Temam, Y. M. Treve,
  \textit{Asymptotic analysis of the Navier-Stokes equations},
  Phys. D, \textbf{9} (1983), 157-188.


\bibitem{FP67}
  C. Foias, G. Prodi,
  \textit{Sur le comportement global des solutions non stationnaires des
\'{e}quations de Navier-Stokes en dimension two},
  Rend. Sem. Mat. Univ. Padova, \textbf{39} (1967), 1-34.


\bibitem{FSTe88}
  C. Foias, G. R. Sell, R. Temam,
  \textit{Inertial manifolds for nonlinear
           evolutionary equations},
  J. Differential Equations, \textbf{73} (1988), 309-353.


\bibitem{FST89}
  C. Foias, G. R. Sell, E. Titi,
  \textit{Exponential tracking and approximation of inertial
           manifolds for dissipative nonlinear equations},
  J. Dynamics and Differential Equations, \textbf{1 (2)} (1989), 199-244.


\bibitem{FTe79}
  C. Foias, R. Temam,
  \textit{Some analytic and geometric properties of the
solutions of the evolution Navier-Stokes equations},
 J. Math. Pure Appl., \textbf{58} (1979), 339-368.


\bibitem{FT89_Gevrey}
  C. Foias, R. Temam,
  \textit{Gevrey class regularity for the solutions of the Navier-Stokes
equations},
  J. Funct. Anal., \textbf{87} (1989), 359-369.


\bibitem{FT98}
  A. B. Ferrari, E. S. Titi,
  \textit{Gevrey regularity for nonlinear analytic parabolic equations},
  Comm. in Partial Diff. Eq., \textbf{23 (1\&2)} (1998), 1-16.


\bibitem{IT05_Attractor}
  A. A. Ilyin and E.S. Titi,
  \textit{Sharp estimates for the number of degrees of freedom for the damped-
driven 2D Navier--Stokes}, Journal of Nonlinear Science, \textbf{16}
(3) (2006), 233-253.


\bibitem{JT92}
  D. A. Jones, E. S. Titi,
  \textit{Determination of the solutions of the Navier-Stokes equations by
finite volume elements},
  Phys. D, \textbf{60} (1992), 165-174.


\bibitem{JT93_Detmodesnodes}
  D. A. Jones, E. S. Titi,
  \textit{Upper bounds on the number of determining modes, nodes, and volume
elements for the Navier-Stokes equations},
  Indiana University Mathematics Journal, \textbf{42} (1993), 875-887.


\bibitem{LL77_Hydrodynamics}
  L. D. Landau, E. M. Lifshitz,
  \textit{Fluid Mechanics},
  Pergamon, Oxford 1977.


\bibitem{LP98}
  V. S. L'vov, E. Podivilov, A. Pomyalov, I. Procaccia, D. Vandembroucq,
  \textit{Improved shell model of turbulence},
  Physical Review E. \textbf{58 (2)} (1998), 1811-1822.


\bibitem{LPP99}
  V. S. L'vov, E. Podivilov, I. Procaccia,
  \textit{Hamiltonian structure of the Sabra shell model of turbulence:
           exact calculation of an anomalous scaling exponent},
  Europhysics Lett., \textbf{46 (5)} (1999), 609-612.


\bibitem{LM72}
  J. L. Lions, E. Magenes,
  \textit{Non-homogeneous Boundary Value Problems and Applications},
  Springer, Berlin (1972).


\bibitem{MB02}
  A. J. Majda, A. L. Bertozzi,
  \textit{Vorticity and Incompressible Flow},
  Cambridge University press, 2002.


\bibitem{Ma86}
  C. Marchioro,
  \textit{An example of absence of turbulence for any Reynolds number},
  Comm. Math. Phys., \textbf{105} (1986), 99-106.


\bibitem{OY89}
 K.  Okhitani, M. Yamada,
  \textit{Temporal intermittency in the energy cascade process and
           local Lyapunov analysis in fully developed model of turbulence},
  Prog. Theor. Phys., \textbf{89} (1989), 329-341.


\bibitem{Sri-Ou}
  Y.-R. Ou and S.S. Sritharan,
  \textit {Analysis of regularized Navier-Stokes equations. I, II.}
  Quart. Appl. Math., \textbf {49}  (1991),  651-685, 687-728.


\bibitem{SchKaLo95}
  N. Schr\"{o}ghofer, L. Kadanoff, D. Lohse,
  \textit{How the viscous subrange determines inertial range properties in turbulence shell models},
  Physica D, \textbf{88} (1995), 40-54.


\bibitem{SellYou92_Inertial}
  G.R. Sell, Y. You,
  \textit{Inertial manifolds: the non-self-adjoint case},
  J. Diff. Eq., \textbf{96} (1992), 203-255.


\bibitem{Te84}
  R. Temam,
  \textit{Navier-Stokes Equations: Theory and Numerical Analysis},
  North-Holland, Amsterdam, 1984.


\bibitem{Te88}
  R. Temam,
  \textit{Infinite-Dimensional Dynamical Systems
            in Mechanics and Physics},
  Springer-Verlag, New-York, 1988.


\end{thebibliography}
\end{document}